\documentclass[aps,prl,twocolumn,amsthm,amsmath,amssymb, superscriptaddress,reprint]{revtex4-2}
\usepackage{graphicx}% Include figure files
\graphicspath{{Figures/}}
\usepackage{subfigure}
\usepackage{epsfig}
\usepackage[normalem]{ulem}
\usepackage{dcolumn}% Align table columns on decimal point
\usepackage{bm}% bold math
\usepackage{hyperref}% add hypertext capabilities
\hypersetup{colorlinks=true, citecolor=blue, urlcolor=blue, linkcolor=blue}
\usepackage{appendix}
\usepackage{mathrsfs}
\usepackage{comment}
\usepackage{hyphenat}

\newcommand{\Veph}{V_{\text{e-ph}}}

\renewcommand{\v}[1]{{\boldsymbol{#1}}}

\newcommand{\imth}{\hspace{1pt}\mathrm{i}\hspace{1pt}}
\newcommand{\abs}[1]{\left\lvert#1\right\rvert}

\newcommand{\s}{{\sigma}}

\def\beq{\begin{equation}}
\def\eeq{\end{equation}}
\def\bald{\begin{aligned}}
\def\eald{\end{aligned}}
\def\bea{\begin{eqnarray}}
\def\eea{\end{eqnarray}}

\def\inc#1{\left(#1\right)}

\def\ket#1{\left|#1\right\rangle}
\def\avg#1{\left\langle#1\right\rangle}

\usepackage[compat=1.1.0]{tikz-feynman}
\newcommand{\plaquette}{\kern.08em
\begin{tikzpicture}[baseline={([yshift=-1ex]current bounding box.center)}]
    \draw coordinate (a) at (0,0);
    \draw coordinate (b) at (0.1,0);
    \draw coordinate (d) at (0.06,0.09);
    \draw coordinate (c) at (0.16,0.09);
    \draw (a) -- (b) -- (c) -- (d) -- (a) pic [draw=black]{} ;
\end{tikzpicture}%
\kern.08em%
}

\newcommand{\plaquettedimerone}{\kern.08em
\begin{tikzpicture}[baseline={([yshift=-.5ex]current bounding box.center)}]
    \draw coordinate (a) at (0,0);
    \draw coordinate (b) at (0.3,0);
    \draw coordinate (d) at (0.18,0.26);
    \draw coordinate (c) at (0.48,0.26);
    \draw[color=black!60] (a) circle (0.04);
    \draw[color=black!60] (b) circle (0.04);
    \draw[color=black!60] (c) circle (0.04);
    \draw[color=black!60] (d) circle (0.04);
    \draw[black,very thin] (a) -- (b);
     \draw[black,very thin] (c) -- (d);
    \draw[black, ultra thick] (b)--(c);
    \draw[black, ultra thick] (d)--(a);
\end{tikzpicture}%
\kern.08em%
}

\newcommand{\plaquettedimertwo}{\kern.08em
\begin{tikzpicture}[baseline={([yshift=-.5ex]current bounding box.center)}]
    \draw coordinate (a) at (0,0);
    \draw coordinate (b) at (0.3,0);
    \draw coordinate (d) at (0.18,0.26);
    \draw coordinate (c) at (0.48,0.26);
    \draw[color=black!60] (a) circle (0.04);
    \draw[color=black!60] (b) circle (0.04);
    \draw[color=black!60] (c) circle (0.04);
    \draw[color=black!60] (d) circle (0.04);
    \draw[black,ultra thick] (a) -- (b);
     \draw[black,ultra thick] (c) -- (d);
    \draw[black, very thin] (b)--(c);
    \draw[black, very thin] (d)--(a);
\end{tikzpicture}%
\kern.08em%
}

\newcommand{\trimerdimer}{\kern.08em
\begin{tikzpicture}[baseline={([yshift=-.5ex]current bounding box.center)}]
    \draw coordinate (a) at (-0.2,0.2);
    \draw coordinate (b) at (0,0);
    \draw coordinate (c) at (0.2,0.2);
    \draw coordinate (d) at (0.4,0);
    \draw coordinate (e) at (0.6,0.2);
    \draw[color=black!60] (a) circle (0.04);
    \draw[color=black!60] (b) circle (0.04);
    \draw[color=black!60] (c) circle (0.04);
    \draw[color=black!60] (d) circle (0.04);
    \draw[color=black!60] (e) circle (0.04);
    \draw[black,double,thin] (a) -- (b);
    \draw[black,double,thin] (b) -- (c);
    \draw[black,very thin] (c) -- (d);
    \draw[black, ultra thick] (d)--(e);
\end{tikzpicture}%
\kern.08em%
}

\newcommand{\dimertrimer}{\kern.08em
\begin{tikzpicture}[baseline={([yshift=-.5ex]current bounding box.center)}]
    \draw coordinate (a) at (-0.2,0.2);
    \draw coordinate (b) at (0,0);
    \draw coordinate (c) at (0.2,0.2);
    \draw coordinate (d) at (0.4,0);
    \draw coordinate (e) at (0.6,0.2);
    \draw[color=black!60] (a) circle (0.04);
    \draw[color=black!60] (b) circle (0.04);
    \draw[color=black!60] (c) circle (0.04);
    \draw[color=black!60] (d) circle (0.04);
    \draw[color=black!60] (e) circle (0.04);
    \draw[black,double,thin] (d) -- (e);
    \draw[black,double,thin] (d) -- (c);
    \draw[black,very thin] (c) -- (b);
    \draw[black, ultra thick] (a)--(b);
\end{tikzpicture}%
\kern.08em%
}

\def\Eq#1{Eq.~(\ref{#1})}
\def\Fig#1{Fig.~\ref{#1}}

\begin{document}
\title{
Quantum spin liquid 
from 
electron-phonon coupling
}

\author{Xun Cai}
\affiliation{Beijing National Laboratory for Condensed Matter Physics and Institute of Physics, Chinese Academy of Sciences, Beijing 100190, China}

\author{Zhaoyu Han}
\affiliation{Department of Physics, Stanford University, Stanford, CA 94305, USA}

\author{Zi-Xiang Li}
\email{zixiangli@iphy.ac.cn}
\affiliation{Beijing National Laboratory for Condensed Matter Physics and Institute of Physics, Chinese Academy of Sciences, Beijing 100190, China}

\author{Steven A. Kivelson}
\email{kivelson@stanford.edu}
\affiliation{Department of Physics, Stanford University, Stanford, CA 94305, USA}

\author{Hong Yao}
\email{yaohong@tsinghua.edu.cn}
\affiliation{Institute for Advanced Study, Tsinghua University, Beijing 100084, China}

\begin{abstract}
A quantum spin liquid (QSL) is an exotic insulating phase with emergent gauge fields and fractionalized excitations. However, the {\it unambiguous} demonstration of the existence of a QSL  in 
 a ``non-engineered'' microscopic model (or in any  material) remains challenging.
Here, using \textit{numerically-exact} sign-problem-free quantum Monte Carlo simulations, we 
show that a QSL 
arises in a  non-engineered electron-phonon model. 
Specifically, we investigate the ground-state phase diagram of the bond Su-Schrieffer-Heeger (SSH) 
model on a 2D triangular lattice 
at half filling (one electron per site)
which we show includes a  QSL phase which is fully gapped, exhibits 
no symmetry-breaking order, 
and supports  deconfined fractionalized holon excitations. This suggests new routes for finding QSLs in realistic materials and high-$T_c$ superconductivity by lightly doping them.

\end{abstract}
\date{\today}

\maketitle
{\bf Introduction.} One of the central focuses of modern  condensed matter physics has been the quest for a physical system with a quantum spin liquid (QSL) ground state \cite{Anderson1973}, an exotic insulating state harboring fractionalized excitations, deconfined emergent gauge fields, 
and topological order. (For reviews, see, e.g. Refs. \cite{PALee2008ScienceReview, Balents2010Review, Savary2016, Norman2016RMP, XGWen-RMP2017, Zhou2017RMP,Knolle-Moessner2019, KivelsonReviewQSL,Kivelson2023NatureReview}.)
Research on QSLs has been largely motivated by their intriguing properties such as fractionalized excitations, potential relevance to high-$T_c$ superconductivity (SC) \cite{Anderson1987Science, KRS1987PRB, Lee2006rmp, Anderson2004vanilla}, and 
their potential applications in topological quantum computation \cite{Kitaev2003, Sarma2008RMP}.
Despite tremendous experimental efforts in 
past decades (see, e.g., Refs. \cite{Takagi2007PRLQSL, Matsuda2010ScienceQSL, Saito2011Nature, Han2012Nature, Lee2015ScienceQSL, Zhao2016NatureQSL, Nagler2016NPRuCl3, Nagler2017ScienceRuCl3,Matsuda2018Nature,Dressel2021Science,Scheie2024proximate}), 
{\it unambiguous} experimental evidence establishing the existence of 
a QSL in any real material remains elusive.
Consequently, analytically or numerically-exact solutions
of  ``natural'' or ``non-engineered'' microscopic models that 
establish the existence of a QSL phase are  of fundamental importance and can  potentially provide useful guidance 
in searching for QSLs in realizable materials.

\begin{figure}[t]
\includegraphics[width= 0.85\linewidth]{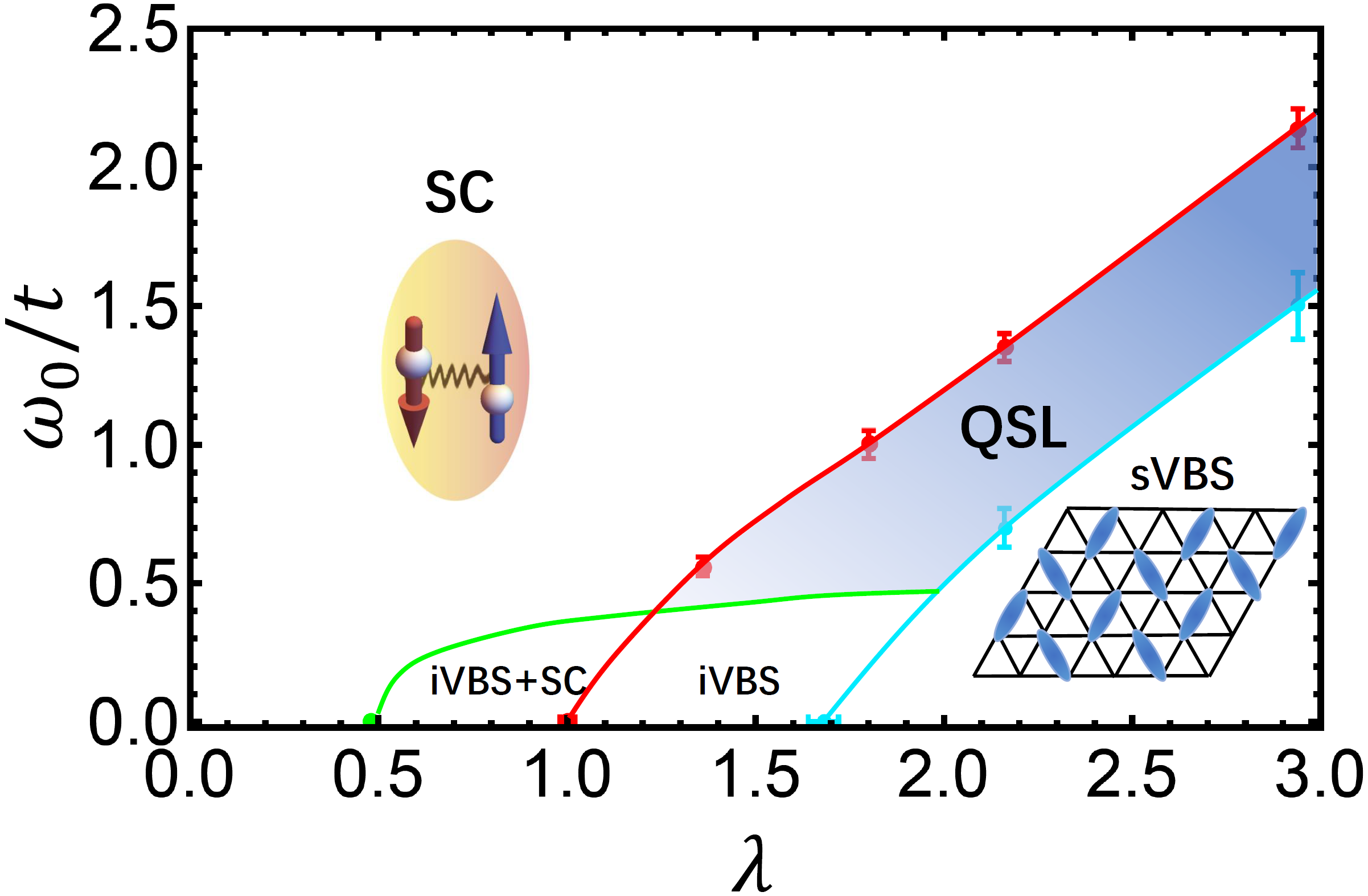}~
\caption{Zero-temperature phase diagram of the SSH electron-phonon model at half-filling on triangular lattice with varying EPC strength $\lambda$ and phonon frequency $\omega_0$, obtained 
from state-of-the-art QMC simulations. Here SC, sVBS, iVBS, and QSL denote superconducting, staggered VBS, incommensurate VBS, and quantum spin liquid, respectively. The QSL 
to SC transition is shown to be continuous and consistent with XY* universality, while the transition to the sVBS is  first order over at least a portion of its extent.
}
\label{Fig1}
\end{figure}

To date, most analytical and numerical studies have been devoted to searching for QSLs in frustrated quantum magnets where only local repulsion between electrons are considered (see e.g., Refs. \cite{Kalmeyer-Laughlin1987, Baskaran1987, Kivelson1988PRL, Sachdev1991PRL,XGWen1991PRB, Moessner2001PRL, Balents2002PRBBFG, XGWen2003PRL, Hermele2004, FaWang2006PRB,  Kitaev2006, Yao2007PRL,YingRan2007, Jackeli-Khaliullin2009,  Yao2011PRL, Melko2011NP,  White2011ScienceKagome, Corboz2012PRXQSL, Mcculloch2012PRL, Jiang2012PRBQSL, Jiang2012NPQSL, Yao2012PRLdimer, Verstraete2013PRLQSL, CXu2013PRL,   He2014PRLQSL, Kee2014PRL, Sheng2015PRBkagome, Xiang2017PRLspinliquid, Pollmann2017PRXDirac, Zhang2018PRL, Zhu2018SA, Song2019NCQSL, He2019PRLDirac, aaron2020prx, Li2022PRB, Zhou2022SB, Gu2022PRXQSL, Jiang2023PRBQSL, Li2024PRL}). Since electron-phonon coupling (EPC) is also ubiquitous in materials, it is natural to ask~\cite{KRS1987PRB} if EPC can be the primary microscopic mechanism of QSL formation. 
Nonetheless, exploration of QSLs induced by pure EPC has been rare partly because EPC 
generates attractions which tend to favor pairing between itinerant electrons instead of spin interactions between local magnetic moments.
A recent development is the establishment of resonating-valence-bond (RVB) states from strong-coupling analyses of two special electron-phonon models defined on a (generalized) Lieb lattice \cite{Kivelson2023PRL,han2024emergent}. 
Moreover, it was recently shown that 
a Su-Schrieffer-Heeger (SSH) type 
EPC on the {\it bipartite} square lattice can induce antiferromagnetic (AF) ordering as well as valence bond solid (VBS) phases \cite{XCai2021PRL, Cai2022PRB, Assaad2022PRBSSH, Scalettar2022PRB,Assaad2024PRB}, which partly motivates us to ask whether such type of EPC can induce a QSL on a {\it non-bipartite} lattice.

Here we show that a QSL can indeed be found in such a non-engineered EPC model on the triangular lattice in a broad, intermediate coupling parameter regime.
Specifically, we investigate a prototypical microscopic model featuring an SSH-type coupling between optical phonons and electrons on a triangular lattice.
This model is sign-problem-free \cite{Troyer2005PRL, CJWu2005PRB,Berg2012Science, ZXLi2015PRB,ZXLi2016PRL,TXiang2016PRL}, so 
we have been able to study it using large-scale  numerically-exact 
determinant quantum Monte Carlo (QMC) simulations \cite{BSS1981PRD, hirsch1985prb, Assaadnote, Werner2016book}.  
The ground state phase diagram  
inferred from the state-of-the-art QMC is shown in Fig.~\ref{Fig1}: The ground state possesses 
SC long-range order at weak EPC or whenever the phonon frequency is sufficiently high. 
 Valence bond solid (VBS) order is dominant for intermediate to strong EPC whenever the 
  phonon frequency is sufficiently small. 

Most significantly, a gapped QSL phase emerges 
  between the SC and VBS phases, as shown by our QMC simulations. 
  Several features accessible to large-scale QMC simulations are used to establish  this:  1) We show that the QSL has a finite gap.  2) 
 There are no spontaneously broken symmetries - at least none with an order parameter magnitude large enough to be detectable.  3)  There exists
  a deconfined (fractionalized) holon excitation 
  that has 
  charge $e$ and no spin. (Modulo only the possibility that we could be fooled were the confinement scale larger than our system sizes, this constitutes smoking-gun evidence~\cite{Kivelson1988PRL} that the indicated phase is a QSL.)  4)  We observe anomalous power-law correlations at the apparently continuous quantum phase transition between the QSL and the SC phases which is consistent with the XY$^*$ universality, expected~\cite{Senthil-Fisher2004PRB} in this circumstance.  We moreover provide a heuristic understanding of the emergence of QSL phases in this model employing a strong-coupling perspective. To the best of our knowledge, this is the first time that a QSL has been shown to emerge in a non-engineered EPC model.

The present results should serve as an inspiration to substantially broaden the search for QSL candidate materials - to include materials without any obvious magnetism but with strong EPC involving relatively high frequency phonons, especially in cases in which the electronic bands are relatively flat.  The proximity of the QSL to a SC phase 
is also suggestive that high-$T_c$ SC might occur in proximity to such a QSL phase.

{\bf Model.} We consider the bond SSH  
model on a triangular lattice described by the Hamiltonian:
\bea\label{Hamiltonian}
&&\hat H=
\sum_{\avg{ij}}
\Big[-\inc{t+ g\hat X_{ij}}
\hat B_{ij}
+\frac{\hat{P}_{ij}^2}{2M}+\frac{K}{2}\hat{X}_{ij}^2 \Big],
\eea
where $\hat B_{ij}=\sum_\s(c^\dagger_{i\s} c_{j\s} + \mathrm{h.c.})$ is the bond-density operator in which $c^\dagger_{i\s}$ creates an electron on site $i$ with spin polarization $\s=\uparrow,\downarrow$, 
and $\hat{X}_{ij}$ and $\hat{P}_{ij}$ are, respectively, the displacement and momentum operators of an optical phonon residing on the bond between nearest-neighbor (NN) sites $\avg{ij}$.  Here $t$ is the bare nearest-neighbor (NN) hopping amplitude, and
$g$ is the strength of the EPC. 
The relevant dimensionless 
measure of the EPC strength is $\lambda \equiv
\frac{zg^2}{KW}$
where $z=6$ is the number of nearest-neighbors and $W=9t$ is the electron's bare bandwidth on triangular lattice~\cite{Ly2023arXiv}.
For simplicity we have assumed Einstein phonons with bare frequency $\omega_0 = \sqrt{K/M}$.
We set $\hbar=1$ throughout this paper.

Because this model is free from the notorious fermion minus sign problem,  we have been able to perform numerically-exact QMC simulations to access 
the ground-state (zero-temperature) properties 
up to system sizes $18\times 18$. To do this, we implemented  projective determinant QMC simulations \cite{Assaad2001PRB}. We have focused on the model 
at half filling by fixing  the electron number 
rather than the chemical potential. The DQMC simulations on the model with strong electron-phonon coupling are typically computationally more demanding  than purely electronic models due to the longer auto-correlation times.  We have adopted various strategies to reduce the auto-correlation time; nonetheless heavy computational resources have 
still been required to achieve reliable results with high accuracy (details are in the SM). 

\begin{figure}[t]
\includegraphics[width= 0.95\linewidth]{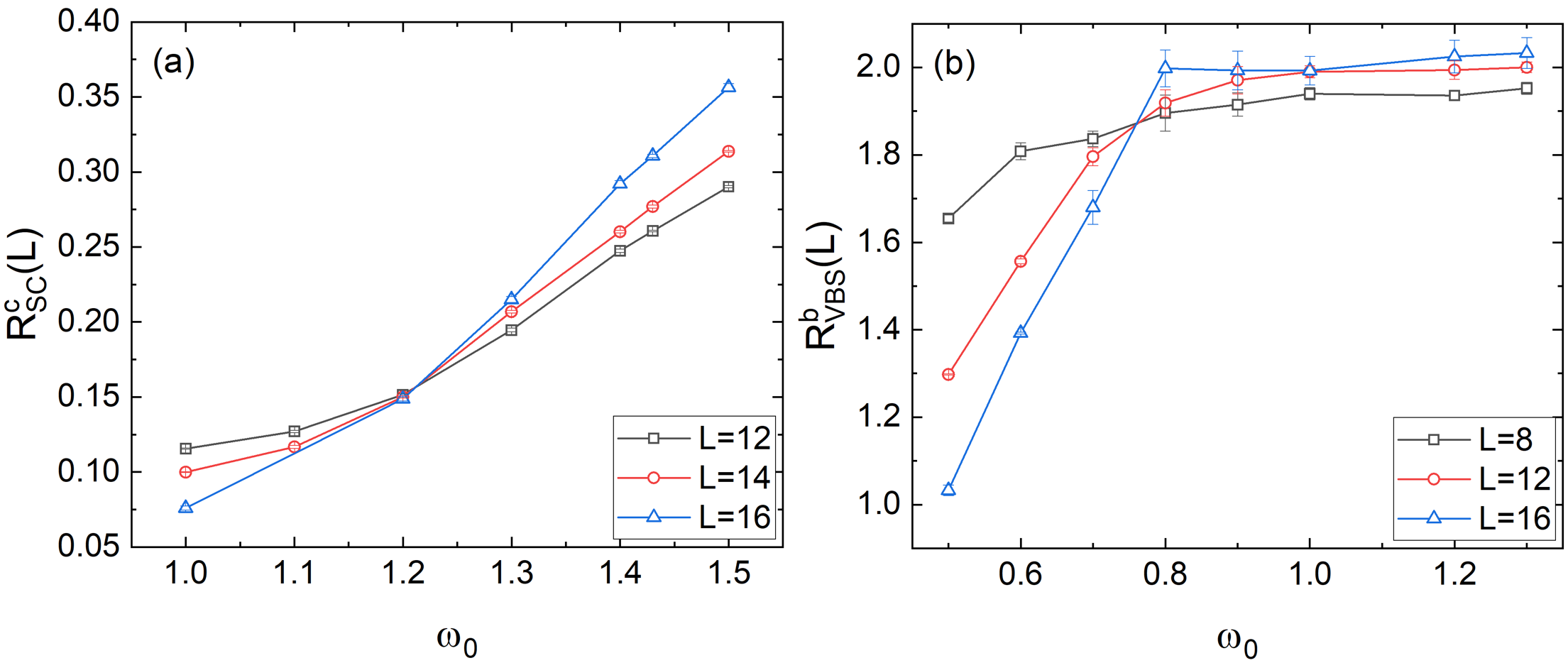}~
\caption{QMC results on the triangular lattice with size $L\times L$. (a) The correlation-length ratio for superconducting order as a function of $\omega_0$ with fixed $\lambda=2.16$. The crossing point for different system sizes indicates the transition point between SC disordered and ordered phases occurs at $\omega_0 \approx 1.2$. (b) The Binder ratio for sVBS order as a function of $\omega_0$ with fixed $\lambda=2.16$. The transition 
between sVBS disordered and ordered phases occurs at $\omega_0 \approx 0.7$.  }
\label{Fig2}
\end{figure}

{\bf Ground-state phase diagram.} The ground-state phase diagram in the extreme adiabatic limit, $\omega_0=0$, is shown along the bottom line of \Fig{Fig1}.  
In this limit, quantum fluctuations of the phonon fields vanish and the phonon configurations $\{X_{ij}\}$ are static, so ground-state properties can be accessed straightforwardly by finding the phonon configurations $\{X_{ij}\}$ that minimize the adiabatic energy.  For small $\lambda$, because the Fermi surface is not perfectly nested, a symmetry-preserving metallic ground-state arises for $\lambda$ smaller than a finite critical value. 
For stronger coupling, 
incommensurate VBS (iVBS) long-range order is found for $ \lambda_{c1} < \lambda < \lambda_{c2} $
($\lambda_{c1} \approx 0.5$ and $\lambda_{c2} \approx 1.7$) with a $\lambda$ dependent ordering vector that is not far from the optimal Fermi surface nesting vector.  
Even in the presence of iVBS order, ungapped  
pockets of Fermi surface persist for $\lambda_{c1}<\lambda < \lambda_{FS}$ ($\lambda_{FS}\approx 1.0$), while the fermionic spectrum is fully gapped for $\lambda_{FS} < \lambda<\lambda_{c2}$. 
At stronger coupling, $\lambda >
\lambda_{c2}$, commensurate VBS 
order emerges with the staggered pattern shown in the inset of \Fig{Fig1}.
The transitions at $\lambda_{c1}$ and $\lambda_{FS}$ are continuous while the transition at $\lambda_{c2}$ is first order, accompanied by a discontinuous jump in the ordering vector (see details in the SM).

\begin{figure*}[th]
\includegraphics[width= 0.99\linewidth]{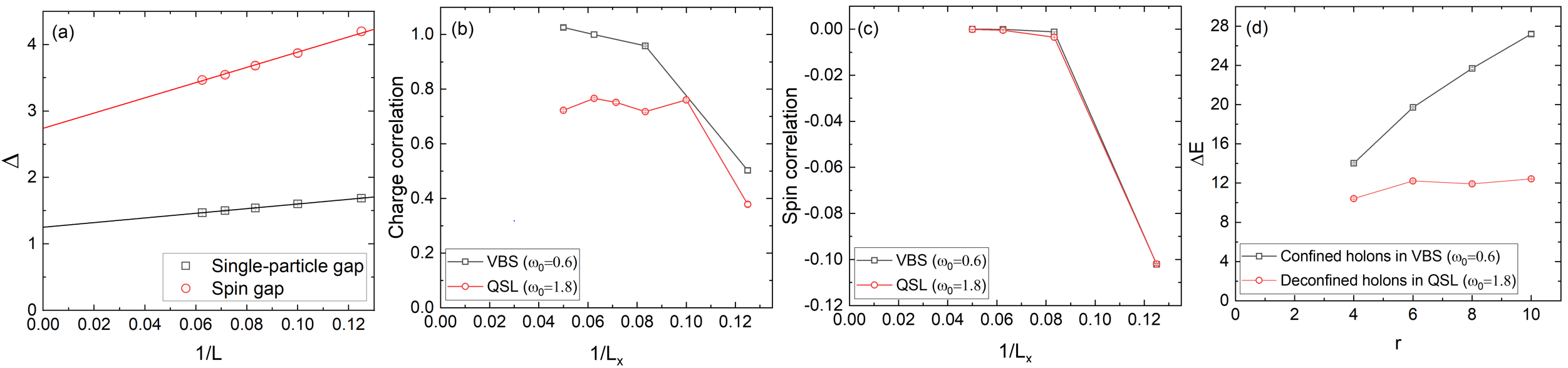}~
\caption{(a) Finite-size-scaling results of single-particle gap and spin gap in the QSL phase for $\lambda=2.16$ and $\omega_0 = 1.0$. Both single-particle and spin gaps are finite, confirming the nature of a gapped Mott insulating phase in the QSL regime. 
(b-d) QMC results for two holes doped away from half filling. We add large on-site impurity potential $V=30$ on two separated sites in the lattice 
to trap the doped holes. The EPC strength $\lambda=2.94$ and two representative values of $\omega_0$ (VBS for $\omega_0=0.6$ and QSL for $\omega_0=1.8$) are considered. The system size in the simulation is $L_x \times 8$. The following observable is considered in the simulation: (b) The correlation of hole density in the regimes around the two impurity potentials. (c) The correlation of $S^z$ in the regimes around the two impurity potentials. (d) The energy difference between the state of $N-2$ electrons with and without large impurity potentials.}
\label{Fig3}
\end{figure*}

The phase diagram for small but non-zero $\omega_0$ is readily inferred by continuity from  $\omega_0=0$, as we have verified for finite $\omega_0$ using QMC. Specifically, VBS persists  
to non-zero $\omega_0$.  Moreover,  with or without VBS order, the presence of a Fermi surface at $\omega_0=0$ for all $\lambda < \lambda_{FS}$ implies the existence of a Cooper instability and hence the emergence of superconducting order at small but non-zero $\omega_0$ - an expectation that 
is consistent with our QMC results in the 
studied range $\omega_0/t \gtrsim 0.5 $. 
In the regime close to the onset of iVBS order, 
SC should coexist with iVBS. Generally, the SC phase expands and the VBS phases are pushed to higher $\lambda$ with increasing $\omega_0$.
Most significantly, in an intermediate regime of 
$\omega_0$ and $\lambda$, we find a fully gapped Z$_2$ QSL in the ground state.

{\bf Broken symmetries.} To determine where in the phase diagram the ground-state has various patterns of spontaneous symmetry-breaking, we have 
computed the structure factor $S(\v{q},L)=\frac{1}{L^4}\sum_{ij}e^{i\v{q}\cdot (\v{r}_i-\v{r}_j)}\langle \hat O^\dagger_i \hat O_j\rangle $ and the associated correlation-length ratio $R^c(L) = 1-\frac{S(\v{Q}+\delta \v{q},L)}{S(\v{Q},L)}$, where $\hat O_j$ is one of a number of possible local order-parameter fields, $\v{Q}$ is the momentum at which the structure factor peaks and $|\delta \v{q}|=4\pi/\sqrt{3}L $ is the minimum crystalline momentum on a lattice with system size $L$$\times$$L$. In a broken symmetry phase, $S(\v{Q},L) \to |\Phi|^2$ and $R^c(L) \to 1$ as $L\to \infty$, where $|\Phi|$ is the magnitude of the order parameter, while in a phase which preserves the requisite symmetries such that $\Phi=0$, it follows that $S(\v{Q},L) \to 0$ and $R^c(L) \to 0$ as $L\to \infty$.  As a function of a control parameter, a symmetry-breaking transition can be identified, as in \Fig{Fig2}(a), with a crossing point at which $R^c(L)$ transitions from being a decreasing to an increasing function of $L$, i.e. a point at which $R^c(L)$ is independent of $L$ for large $L$.
We also consider the Binder ratio $R^b(L)$ to identify the phase transition to a state with 
spontaneous symmetry breaking, as shown in \Fig{Fig2}(b). Details of correlation-length and Binder ratios are in the SM.

An example of this analysis 
is shown in \Fig{Fig2}, where 
for a fixed value of $\lambda=2.16$ (corresponding to $g=1.8$) we have computed the superconductivity correlation-length ratio $R^c_{\rm SC}$ (i.e. with $\hat O_i$ corresponding to a pair-field creation operator $c_{i\uparrow}c_{i\downarrow}$ and $\v{Q}=\v{0}$) and the staggered VBS Binder ratio $R^b_{\rm VBS}$ (i.e. with $O_i$ corresponding to a valence bond density  operator and $\v{Q}=(\pi,0)$) as a function of
$\omega_0$ for $L$ up to $L=16$.
The SC correlation-length ratio $R^c_{\rm SC}(L)$ (shown in \Fig{Fig2}(a)) increases with system size when $\omega_0 > 1.2$, indicating that the ground state possesses long-ranged SC order, and decreases for $\omega_0 < 1.2$, implying only short-range SC correlations. At sufficiently low frequency $\omega_0<0.7$, the order in the ground state 
is a staggered VBS, as evidenced by the results of VBS Binder ratios (shown in \Fig{Fig2}(b)). 
Consequently, in the intermediate frequency regime $0.7<\omega_0<1.2$, the ground state 
has neither SC nor VBS order.  
We have repeated the same analysis for $\lambda=2.94$ (
i.e. $g=2.1$), 
and again found an intermediate range of $\omega_0$ with neither SC nor VBS ordering, as shown in \Fig{Fig1}. We further checked 
for and found no evidence indicating other possible spontaneously broken symmetries  in the intermediate phase, including other forms  VBS ordering (other values of $\v{Q}$) such as columnar VBS and $\sqrt{12}$$\times$$\sqrt{12}$ plaquette VBS, and loop current ordering (details are included in SM).  There thus apparently exists, in a range of intermediate $\lambda$ and $\omega_0$, a symmetric phase without any spontaneous symmetry breaking, 
consistent with a QSL phase. 

{\bf Spectral features of the QSL.}  We 
have also obtained information concerning the single-particle and neutral spin spectra by analyzing the corresponding imaginary-time correlators in the intermediate phase with no symmetry breaking order.  In both cases, we find that the local, two-time correlator falls exponentially with time, implying a gap in the spectrum.  (Details of the fitting procedure are presented in the SM.) The single-particle and spin gap inferred in this way for representative couplings, $\lambda =2.16$ and $\omega_0=1.0$, in the intermediate phase 
are shown for various system sizes, $L$, in \Fig{Fig3}(a), from which it is apparent that they both extrapolate to a finite value in the limit $L\to \infty$. The bond-bond correlations obtained from QMC are short-ranged with a short correlation length (less that one lattice constant);  the singlet gap is smaller but also finite (see the SM for details).
Consequently, the symmetric intermediate phase is fully gapped. According to the celebrated Lieb-Schultz-Mattis-Oshikawa-Hastings (LSMOH) theorem \cite{LSM1961, oshikawa2000prlKondo,Hastings2004PRB}, any symmetric phase with a finite excitation gap 
on a triangular lattice at half filling of spin-$\frac{1}{2}$ electrons cannot be a 
trivial phase; it must be a gapped QSL with accompanying topological order. 

{\bf Fractionalized excitations. 
} To corroborate its existence, 
it is desirable to establish direct signatures of a QSL, for instance the existence of fractionalized excitations such as deconfined holons.
The energy cost of creating two spatially separated holons in a QSL  approaches 
a finite constant even when they are far from each other. 
In contrast, in any topologically-trivial phase such as a VBS, holons are confined in the sense that the energy cost of creating two far separated holons increases linearly with separation. (Since the spin-gap is finite, the energy cost of two far separated charges in a confining phase  eventually saturates at large separation at an energy of order the spin-gap. Moreover,  rather than two far separated holons, in a confining phase the far separated charges consist of two charge e, spin 1/2 ``holes'' - which from this perspective are each viewed as  
bound-states of a holon and a charge 0 spin 1/2 ``spinon.'')

To investigate holon deconfinement we perform QMC simulations for the same model but with two electrons (with opposite $S^z$) removed from the half-filled system.  Moreover, we localize the associated excitations by adding two ``impurity'' potentials  to the Hamiltonian, 
\begin{align}
    \hat H \to \hat H + \sum_{\alpha=1,2}
    V\hat n_{\vec{r}_\alpha} 
\end{align}
which couple to the local charge (relative to half-filling) at two separated sites at $\vec{r}_1$ and $\vec{r}_2$, respectively. 
The simulation have been performed on  lattices of size $L_x\times L_y$, with 
$L_y=8$, and with the
two impurity sites  separated in the $x$ direction by  $r_{12}=|\vec r_1-\vec r_2|=L_x/2$. 
We focus particularly on the energy difference between the doped systems with and without impurity potential $V$, and the two point correlations, $\langle \hat{Q}_1 \hat{Q}_2 \rangle$
and $\langle \hat{\cal S}^z_1 \hat{\cal S}^z_2 \rangle$ where $\hat{Q}_{\alpha} \equiv \sum_{i\in R_\alpha} (\hat{n}_i-1)$   and z-component of spin, $\hat{\cal S}^z_{\alpha} \equiv \sum_{i\in R_\alpha} \hat{S}^z_i$ in each of two spatially separated regions   
$R_{\alpha}$ ($R_\alpha$ consisting of the site $\vec r_\alpha$ and its first and second neighbors) for different values of $L_x$. The on-site impurity potential is fixed at $V=30$.

The results of charge and spin correlations are shown in \Fig{Fig3}(b) and (c), respectively.
In the QSL regime, with increasing $L_x$ the charge correlation grows to a value close to one, while the spin correlation vanishes, implying the excitation trapped by each impurity potential is a holon with charge-$e$ but spin-zero. The energy difference $\Delta E$ between the states of the doped holes with and without impurity potentials, is shown in \Fig{Fig3}(d); $\Delta E$ approximately saturates to a constant  
for $r_{12}\geq 6$, as expected for deconfined holons. For comparison, we performed similar calculations in the regime of VBS.
The behaviors of charge and spin correlations in the VBS phase are qualitatively similar to the ones in the QSL regime, as presented in \Fig{Fig3}(b) and (c), implying the excitations trapped by impurity potentials in the VBS are also holons; 
however, the energy cost $\Delta E$ 
increases roughly linearly with $r_{12}$, as expected in a (weakly) confining phase.
The results are consistent with the expectation of decofined holons only in the QSL phase.

\begin{figure}[t]
\includegraphics[width= 0.85\linewidth]{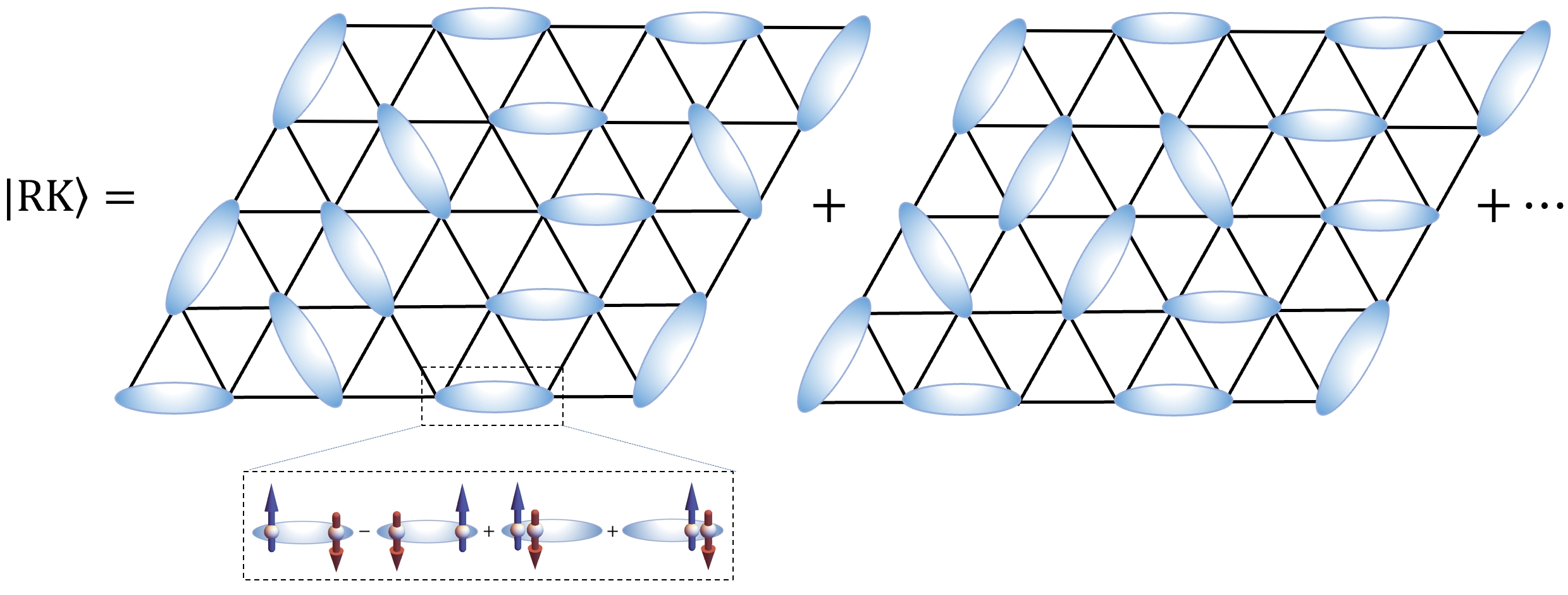}
\caption{A schematic representation of the  
RK wave function on the triangular lattice. 
Each 
dimer represents two electrons occupying the bonding 
orbital between two neighboring sites. }
\label{Fig4}
\end{figure}

{\bf Quantum phase transitions.} The  quantum phase transition between the insulating state 
and the neighboring SC phase  
is expected to exhibit distinct features if the insulating phase supports fractionalized excitations. 
Specifically, the quantum phase transition from a QSL with $Z_2$ topological order 
to a conventional (s-wave) SC 
can proceed through the condensation of holons/doublons (``chargeons'');  the corresponding critical properties~\cite{Senthil-Fisher2004PRB}
are characterized by 3d XY* universality, reflecting the existence of fractionalized excitations and exhibiting power-law correlations at criticality with a very large anomalous dimension, $\eta_{\text{XY*}}\approx 1.49$~\cite{Isakov2012Science}.  By contrast, the transition between a conventional insulator and a SC, or between a QSL and an exotic SC$^*$ phase (with deconfined visons) is expected to exhibit conventional 3d XY criticality, i.e. with $\eta_{XY} \approx 0.034$.
By numerically analyzing the pair-pair correlations at the QSL-SC transition in the present model, as detailed in the SM, we found a $\eta\approx 1.45 \pm 0.08$,  
consistent with 
a XY$^*$ transition within the error bar. 
This further confirms that the symmetry-preserving insulating phase in the triangular SSH model is a Z$_2$ QSL.

{\bf Strong coupling analysis.}  We can get a feeling for why a QSL phase arises and what its character is, by approaching the problem form the theoretically tractable strong coupling limit, where $t\ll g^2/K$ (i.e. $\lambda\gg 1$) and $\omega_0 \ll g^2/K$.

In the adiabatic limit $\omega_0=0$, complete identification of the degenerate ground-states can be obtained straightforwardly from a combination of analytic analysis and numerical solution of large finite size systems;  the electronic Hamiltonian is non-interacting, so all that is required is to minimize the adiabatic-energy with respect to the phonon coordinates, $\{X_{ij}\}$.  Remarkably, to the first order in $t$, there is an extensive number of 
degenerate ground-states,   
which we will refer to as hard-core dimer states. 
They can be 
associated with possible fully packed configurations $\mathcal{C}$ of hard-core dimers, where each dimer labels a nearest-neighbor bond and where exactly one dimer touches each site of the lattice, as shown in Fig. \ref{Fig4}.  In the 
state $|\mathcal{C}\rangle$, two electrons occupy the 
bonding orbital on each dimer-covered bond so that the expectation value of 
$\langle \hat B_{ij}\rangle = 2$ and  $\langle \hat X_{ij}\rangle = 2g/K$; 
on the remaining bonds,
$\langle \hat B_{ij}\rangle = 0$ and $\langle \hat X_{ij}\rangle = 0$.

Restricting attention to the dimer states, we note that they remain degenerate with each other to first order in both $t$ and $\omega_0$~\footnote{
We have also verified the robustness of the nearly degenerate ground state manifold of the dimer states in the presence of various additional terms, e.g. onsite Hubbard repulsion and Holstein-type EPC.}.  To summarize,  ignoring terms of higher than first order in $t$ and $\omega_0$, 
there is a manifold of degenerate ground-states spanned by a set of linearly independent dimer states, $|\mathcal{C}\rangle$, in one-to-one correspondence with the set of hard-core dimer coverings of the lattice.  
Electron wavefunction of each state is given by 
\bea
|\mathcal{C}\rangle_{e} = \prod_{\{ij\}\in {C}}\ \hat b_{\langle ij\rangle }^\dagger\ |\text{vac}\rangle
\eea
where
$\hat b_{\langle ij\rangle }^\dagger \equiv \frac 1 2\prod_\sigma \left( c^{\dagger}_{i\sigma} + c^{\dagger}_{j\sigma}\right)$
is the valence-bond creation operator and $|\text{vac}\rangle$ is the empty state. The phonon wavefunctions are certain 
Gaussian states for the phonon modes with $ \langle X_{ij} \rangle= 2g/K$ on occupied bonds and 
$ \langle X_{ij} \rangle= 0$ on other bonds (details are given in the SM).

Working to higher order in $t$ and $\omega_0$, we could in principle obtain in terms of the parameters of the original model, an effective model that operates in the hard-core dimer subspace.  The form of the resulting effective hard-core quantum dimer Hamiltonian, $\hat H_\text{dimer} = \hat V + \hat J$, is highly constrained, and indeed of precisely the same form as has been analyzed in various earlier studies in which the origin of the effective model is entirely different \cite{Kivelson1988PRL}.
There are two sorts of terms  - interaction terms, $\hat V$, that are diagonal in the dimer basis, and kinetic terms, $\hat J$, that are off diagonal. To low order in the small parameters, both terms are short-range, so for convenience we will explicitly consider only 
those involving pairs of nearest-neighbor dimers, i.e.
\begin{align}
    \hat{V} = V \sum_{\plaquette}\Big[ \left|\plaquettedimerone
    \right\rangle \left\langle \plaquettedimerone \right| + \left|\plaquettedimertwo
    \right\rangle \left\langle \plaquettedimertwo \right| \Big]\ ,
\end{align}
where  the thick lines represent dimers and the summation is over all possible four-sided plaquettes, and
\begin{align}
    \hat{J} = -J \sum_{\plaquette}\Big[ \left|\plaquettedimerone
    \right\rangle \left\langle \plaquettedimertwo \right| + \left|\plaquettedimerone
    \right\rangle \left\langle \plaquettedimertwo \right| \Big]\ .
\end{align}
(Further range terms that are similar in structure occur as well, although they are still smaller than these most local terms in the strong-coupling limit.)

The interaction terms are relatively simple to compute.  For instance, to second order in $t$ and 
first order in $\omega_0$, the interaction can be shown explicitly to be  $V = t^2K/g^2 - \frac{\sqrt{2}-1}{2}\omega_0$
(details of derivations in the SM). 
The kinetic term $J$ necessarily vanishes 
for any $t$ in 
the $\omega_0 \to 0$ limit as they are tunnelling processes - akin to small bipolaron hopping - involving rearrangements of the phonon coordinates. 
For $0<\omega_0 \ll t/\lambda$, where the potential energy still dominates
the kinetic energy,
the dimer repulsion ($V>0$) favors a staggered VBS ground state,  consistent with the results obtained from the large-scale QMC calculations for large values of $\lambda$ and small $\omega_0/t$.
However, even if we restrict attention to the strong coupling limit, the distinct dependences of $J$ and $V$ on $t$ and $\omega_0$ implies that as a function of $\omega_0/t$ one can access the situation in which  $J\approx V$
\footnote{From the estimates of $V$ and $J$ as a function of $t$ and $\omega_0$ in the strong coupling limit, as shown in the Supplementary Material, one would conclude that $J \sim V$  only if $\omega_0\sim t^2 K/g^2$. 
Since the arguments we have presented so far rely not only on  $\lambda\gg 1$,   
and $\omega_0K/g^2 \ll 1$, this requires we extrapolate the strong coupling results to  
intermediate value of $\lambda$ and $\omega_0 K/g^2$ to access the region studied by QMC. For this reason, the strong coupling analysis of the QSL phase just presented must be viewed as heuristic rather than controlled for the original EPC problem.  
},
in which case it has been shown that (on a non-bipartite lattice) there exists  a perturbatively stable, gapped QSL phase with $\mathbb{Z}_2$ topological order 
~\cite{Moessner2001PRL,Yao2012PRLdimer}. 
In particular, at the special
RK point~\cite{Kivelson1988PRL}, 
the ground state can be expressed as an equal superposition of all possible dimer covering configurations:
\bea
|\text{RK}\rangle={\cal N}^{-1/2}\sum_{\mathcal{C}} |\mathcal{C}\rangle,
\eea
where ${\cal N}$ is the number of hard-core dimer states. A schematic representation of $|\text{RK}\rangle$ is shown in Fig. \ref{Fig4}. 
This is a short-range RVB state with 
$\mathbb{Z}_2$ topological order \cite{XGWen1991PRB,Sachdev1991PRL,Moessner2001PRL, Kitaev2003},
that we believe 
is a representative state of the QSL phase we discovered in the present QMC study.

{\bf 
Concluding remarks.} Using state-of-the-art QMC,
we have established the existence of a QSL phase 
in the ground state of 
SSH electron-phonon models on a triangular lattice on the basis of multiple, independent criteria.
To the best of our knowledge, 
this is the first such case for a non-engineered model, at least for one dominated by EPC.
As SSH-type phonons exist 
commonly in 
quantum materials, our study points out a promising direction to search for QSLs in systems, including  
twisted moire systems~\cite{Bernevig2023phonon}, with 
such phonon couplings.

As the SSH electron-phonon model in \Eq{Hamiltonian} is free from sign problems even for finite doping away from half filling, its physical properties can be accurately studied by large-scale QMC simulations over a broader range of conditions, which is currently one of our ongoing efforts.  We believe that this will 
provide a significant way to explore 
the long-standing problem of high-T$_c$ superconductivity 
emerging from a lightly doped QSL \cite{Jiang2021PRLSCQSL, Li2023PRBdopedVBS, Jiang2020PRLdopedQSL, Jiang2021PRL, Jiang2021Review}, as originally advocated soon after the discovery of cuprate superconductors \cite{Anderson1987Science, KRS1987PRB, Lee2006rmp, Anderson2004vanilla}. 

\textit{Acknowledgement}: We would like to thank Eduardo Fradkin, David Huse, Roger Melko, Subir  Sachdev, T. Senthil, and Zhouquan Wan for helpful discussions; ZH and SAK thank Kyung-Su Kim and John Sous for collaborations on related topics. This work was supported in part by NSFC under Grant No. 12347107 (XC, ZXL and HY), by the Xplorer Prize through the New Cornerstone Science Foundation (HY) and by NSF grant No. DMR-2310312 (ZH and SAK).

\bibliography{QSL-bib}

\begin{appendices}
\widetext
\begin{center}
    \section{Supplemental Material}
\end{center}
\setcounter{equation}{0}
\setcounter{figure}{0}
\setcounter{table}{0}
\makeatletter
\renewcommand{\theequation}{S\arabic{equation}}
\renewcommand{\thefigure}{S\arabic{figure}}
\renewcommand{\thesubfigure}{\thefigure(\alph{subfigure})}
\renewcommand{\bibnumfmt}[1]{[S#1]}
%\renewcommand{\citenumfont}[1]{S#1}

%\begin{widetext}
%\section{Supplemental Materials}
\subsection{A. Details of QMC simulations}
We implement the algorithm of projector (i.e. zero-temperature) quantum Monte-Carlo (PQMC) to investigate the ground-state properties of the SSH electron-phonon coupling model described in Eq. (1) of the main text. The scheme of the PQMC is based on the principle that one can compute the numerically-exact ground-state properties of the model via the imaginary-time evolution 
starting from any given trial wave function as long as the trial wave function is not orthogonal to the true ground-state wave function, a condition that is generically obeyed for a quantum many-body Hamiltonian. The  ground-state wave function of a given Hamiltonian is accessed as: $| \psi_{G} \rangle = \lim_{\Theta \rightarrow \infty}e^{-\Theta \hat H}|\psi_{T} \rangle$, where $\Theta$ is the projective parameter, $| \psi_{G} \rangle$ is the (unnomralized) true ground-state wave function of the Hamiltonian under consideration and $|\psi_{T} \rangle$ is the trial wave function. The ground-state expectation value of 
an observable is computed as: 
\bea
\avg{\hat{O}} = \frac{\langle \psi_{G}|\hat O|\psi_{G} \rangle}{\langle \psi_{G}|\psi_{G} \rangle} = \lim_{\Theta \rightarrow \infty} \frac{\langle \psi_{T}| e^{-\frac{\Theta}{2} \hat H} \hat O e^{-\frac{\Theta}{2} \hat H} | \psi_{T} \rangle}{\langle \psi_{T}| e^{-\Theta \hat H} | \psi_{T} \rangle}
\label{PQMC}
\eea
The algorithm is intrinsically \textit{unbiased} against the choice of the trial wave function $\ket{\psi_T}$ as long as 
%%%$\ket{\psi_T}\neq\ket{\psi_G}$
$\ket{\psi_T}$ is not orthogonal to $\ket{\psi_G}$. Practically, a finite but sufficiently large $\Theta$ is chosen to guarantee the convergence of the results for the observables of interest with respect to  increasing value of
changing values $\Theta$. We choose $\Theta = 2L$ in most simulations performed in this work and $\Theta=50$ in the parameter regimes in the vicinity of quantum phase transition points. We have checked the 
insensitivity of the results against further increasing the value of $\Theta$. To evaluate expectation values in \Eq{PQMC}, we implement the Trotter decomposition to discretize $\Theta$ into $N_{\tau}$ small imaginary-time slices $\Delta_{\tau} = \frac{\Theta}{N_{\tau}}$.  The Trotter error arising from discretization scales as $\Delta_{\tau}^2$. In our simulation, we choose $\Delta_{\tau} = 0.1$ and have checked that the results are convergent by comparing with smaller $\Delta_{\tau}$. We perform the standard procedure of orthogonalization in PQMC every 5 time slices to guarantee the numerical stabilization in the matrix multiplication.

For the QMC simulations on the model with strong electron-phonon coupling, the auto-correlation time in the simulation increases rapidly with decreasing temperature, particularly when the phonon frequency is low. Since we study the ground-state properties of the model, the auto-correlation time becomes very long if $\Theta$ is large in the simulation. To reduce the auto-correlation time and improve the efficiency, we combine local updates and global updates in the employment of the Metropolis algorithm. In the global update, we 
update the phonon fields in all the imaginary-time slices at a fixed spatial site. After each space-time sweep for the local update, we perform 4-6 global updates with random choices of spatial site. We have also checked that the choice of different initial states does not affect the final converged results as long as sufficiently many update steps are adopted. Typically, to achieve reliable results with relatively small statistical errors we run 150 independent Markov chains with 50000 space-time sweeps for thermalization and 50000 space-time sweeps for the measurement of observables. For the simulation of systems of size $L=18$, the number of Markov chains is increased to 224.  The computational resources required are considerably heavy for the simulations with large system sizes and small phonon frequency. For instance, for $L=16$, the computational resource usage is usually more than 0.1 million core-hours in supercomputers to achieve the results for a single set of the parameters 
$\lambda$ and $\omega_0$.

\subsection{B. The details of results for $\omega_{0} = 0$}
\begin{figure}[t]
\subfigure{\label{figS1a}\includegraphics[width= 0.31\linewidth]{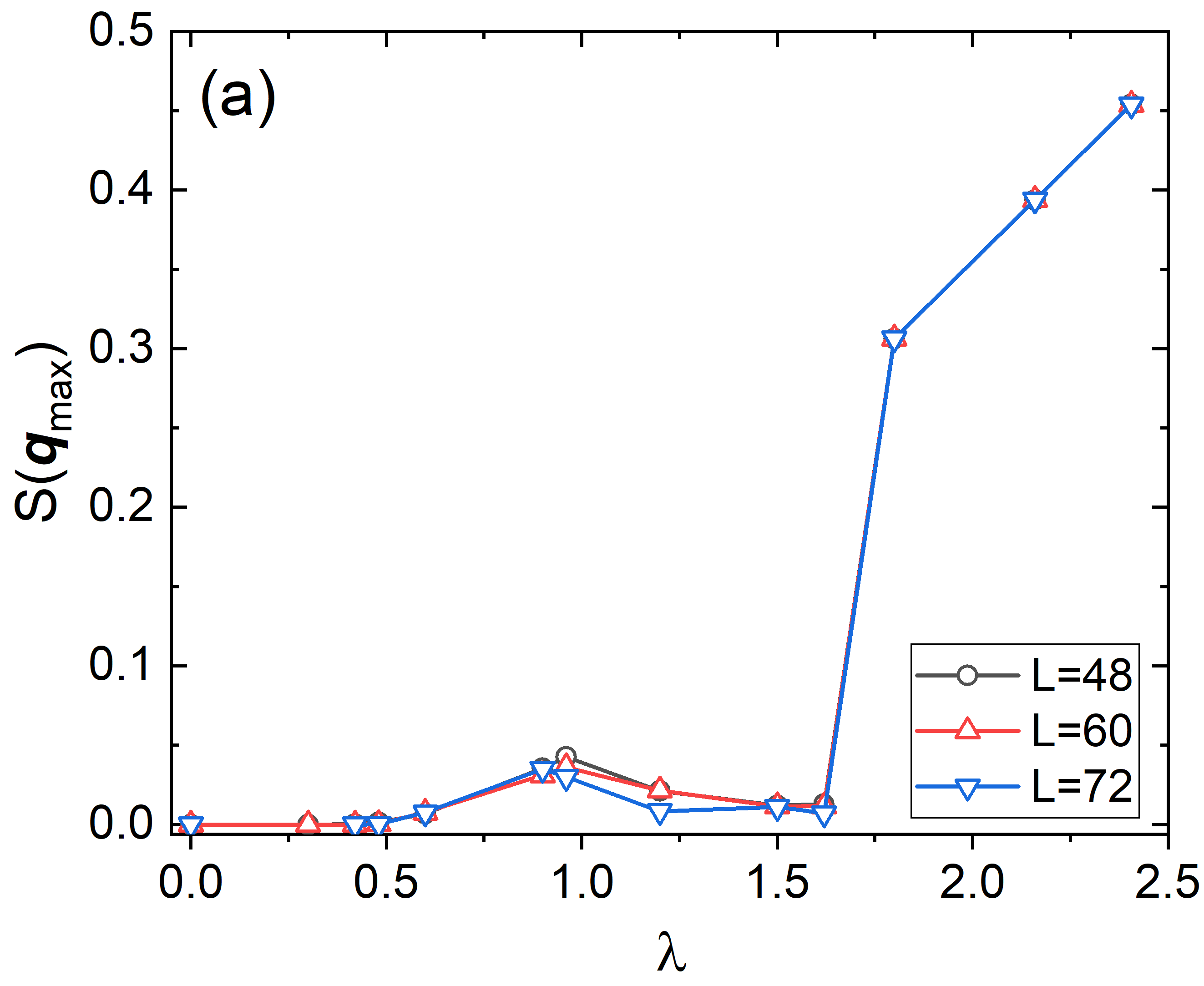}}~~
\subfigure{\label{figS1b}\includegraphics[width= 0.31\linewidth]{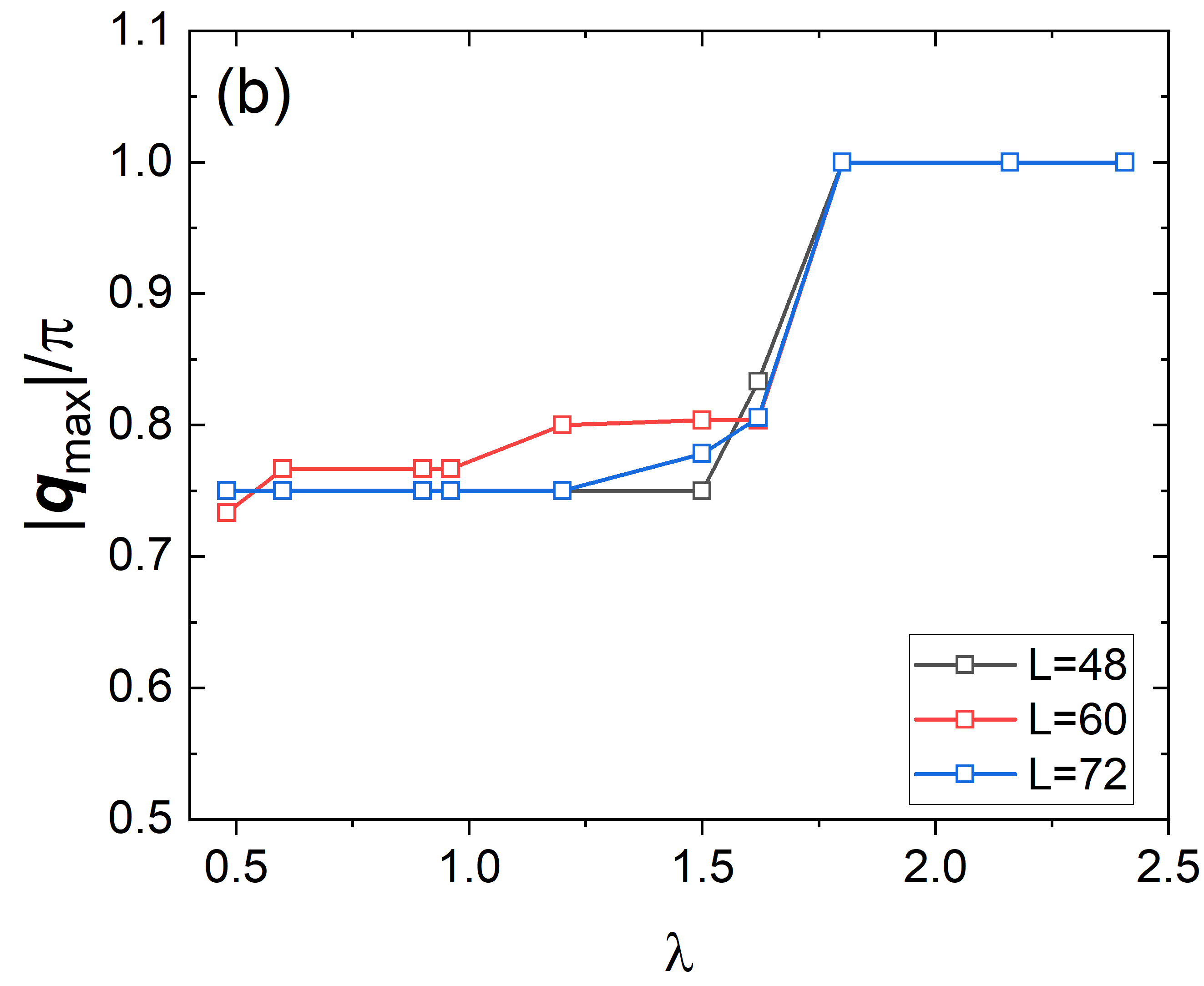}}~~
\subfigure{\label{figS1c}\includegraphics[width= 0.31\linewidth]{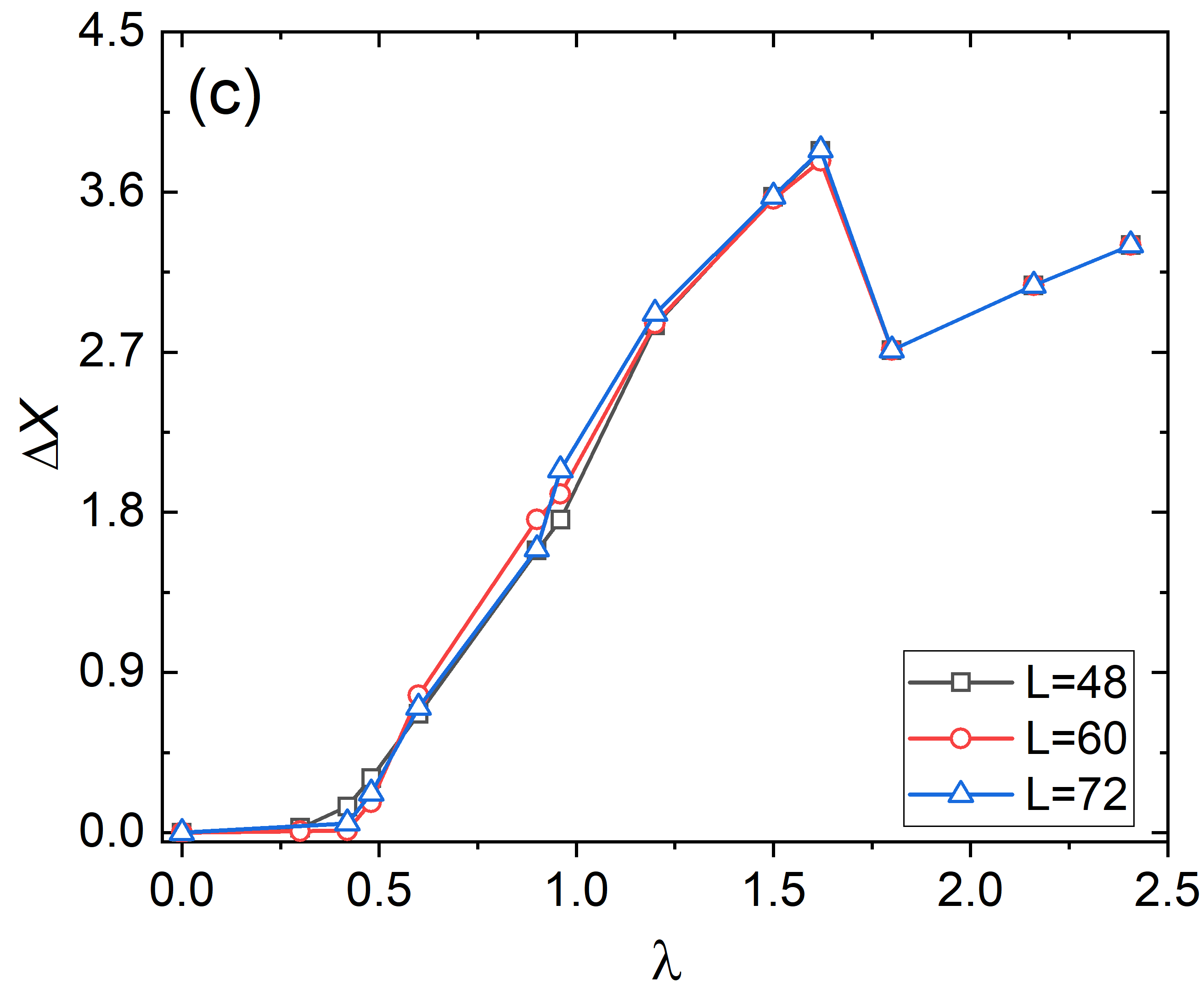}}
\caption{The results for VBS order and structure factor at adiabatic limit $\omega_0=0$. (a) The peak value of VBS structure factor at the corresponding ordering momentum $\v{q}_{\mathrm{max}}$ as the function of $\lambda$. (b) The magnitude of $\v{q}_{\mathrm{max}}$ in multiple of $\pi$ as the function of coupling strength $\lambda$. (c) The maximum difference of phonon displacement $\Delta X$ as the function of $\lambda$. }
\label{FigS1}
\end{figure}

In this section, we present the results for the model in Eq. (1) of the main text in the adiabatic limit, namely $\omega_0 = 0$. In this limit, the kinetic term of phonon is zero and quantum fluctuations of phonon fields vanish. Hence, the phonon displacements $X_{ij}$ are classical variables without quantum dynamics. Upon fixing static phonon displacement configuration $\{X_{ij}\}$, the electrons are described by the quadratic Hamiltonian and the energy of Hamiltonian can be straightforwardly 
computed. Consequently, the ground state of the model can be accessed by minimizing the energy with varying static phonon configuration $\{X_{ij}\}$. In the adiabatic limit, the Hamiltonian of Eq. (1) in the main text is reduced to the following quadratic Hamiltonian depending on static phonon displacement configuration $\{X_{ij}\}$:
\bea
\hat{H}(\{X_{ij}\}) = -\sum_{\avg{ij},\s} (t + g X_{ij}) (c^\dagger_{i\s}c_{j\s} + h.c.) + \frac{K}{2}\sum_{\avg{ij}} X^2_{ij}
\label{meanfield}
\eea
where $t$ is the bare nearest-neighbor hopping amplitude and $g$ is the strength of the SSH EPC. The ground-state energy is minimized through variational method, which yields the self-consistent equation:
\bea
g \avg{\sum_\s c^\dagger_{i\s}c_{j\s}+h.c.} = -K X_{ij}. 
\eea
We solve the self-consistent equation iteratively and reach the phonon displacement configuration $\{X_{ij}\}$ that minimizes the ground-state energy of Hamiltonian in \Eq{meanfield}. 

Without quantum fluctuations, the ground state of \Eq{meanfield} at strong coupling features various dimerized patterns. We summarize the results of VBS order in \Fig{FigS1}. We compute the VBS structure factor for various $\lambda$ and select out the peak value as the magnitude of VBS ordering, as shown in \Fig{FigS1}(a). For sufficiently small electron-phonon coupling strength, the ground state is metallic with no VBS ordering, while VBS long-range order emerges for $\lambda>\lambda_{c1}\approx 0.5$. Besides, the discontinuity around $\lambda=\lambda_{c2}\approx 1.7$ indicates a pronounced first-order transition between two distinct VBS phases. To distinguish between different patterns, we investigate the ordering momentum $\v{q}_{\mathrm{max}}$ identified by the peaked location in momentum space. For $\lambda_{c1}<\lambda<\lambda_{c2}$, the magnitude of the ordering momentum $\abs{\v{q}_{\mathrm{max}}}$ slightly varies with both coupling strength $\lambda$ and system size $L$, featuring an incommensurate VBS (iVBS) order, as shown in \Fig{FigS1}(b). We verify that $\v{q}_{\mathrm{max}}$ is close to the peak of Lindhard function, i.e. the non-interacting response function, especially for relatively weak coupling strength $\lambda<1.0$. It suggests that the properties are still largely influential by the Fermi surface in this regime. Meanwhile, for $\lambda>\lambda_{c2}$, one of the ordering momenta is locked on $\v{q}_{\mathrm{max}}=\inc{\pi,0}$, the wave-vector for staggered VBS (sVBS) pattern. To further corroborate the nature of iVBS order, we calculate the difference between the largest and smallest phonon displacements in $\{X_{ij}\}$, i.e. $\Delta X \equiv X_{\mathrm{max}}-X_{\mathrm{min}}$, as depicted in \Fig{FigS1}(c). For $1.0<\lambda<\lambda_{c2}$, the structure factor decreases while $\Delta X$ increases with $\lambda$, indicating that short-range bond-bond correlation is strong in iVBS state. 

\begin{figure}[tb]
\subfigure{\label{figS2a}\includegraphics[width= 0.3\linewidth]{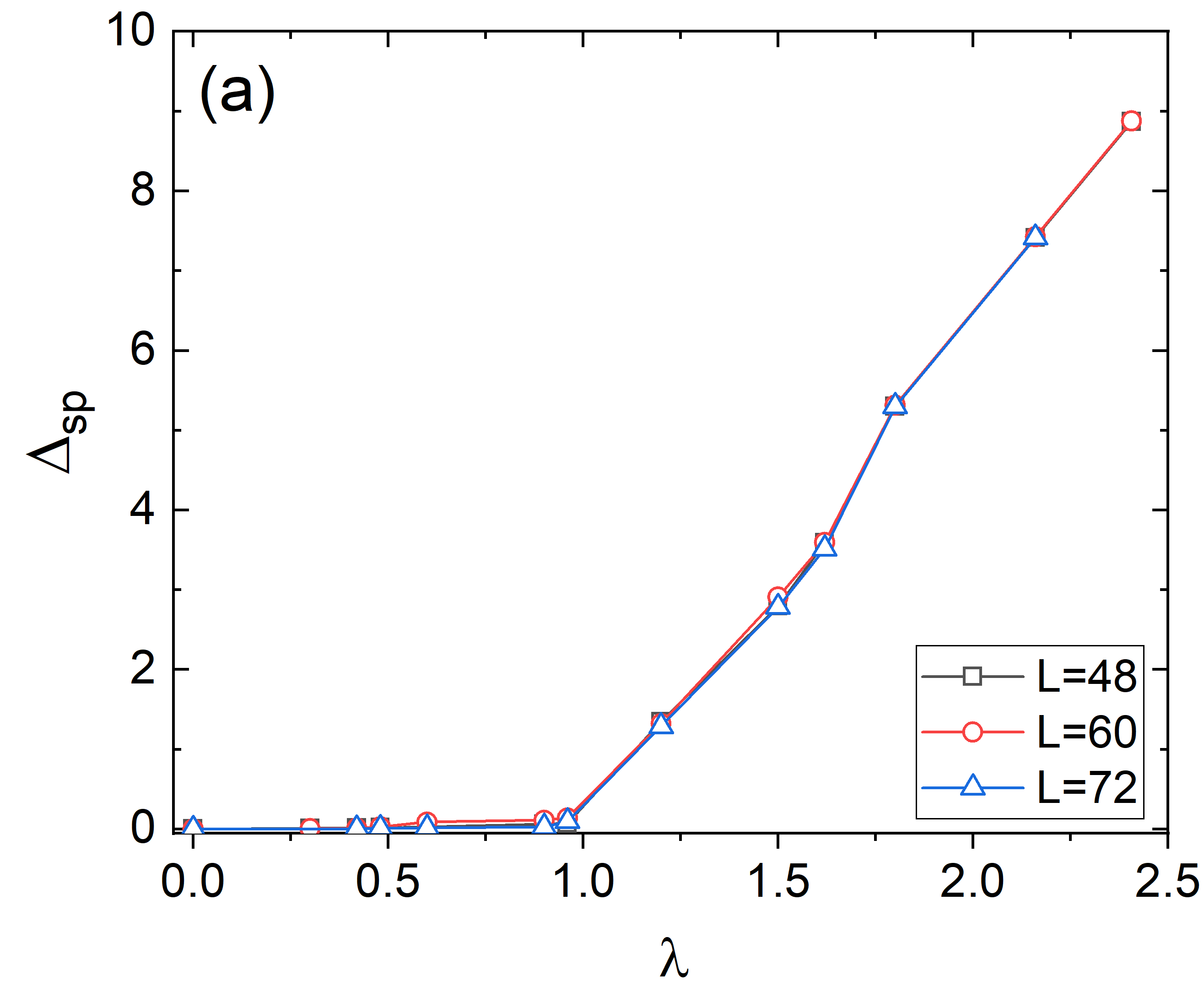}}~~
\subfigure{\label{figS2b}\includegraphics[width= 0.3\linewidth]{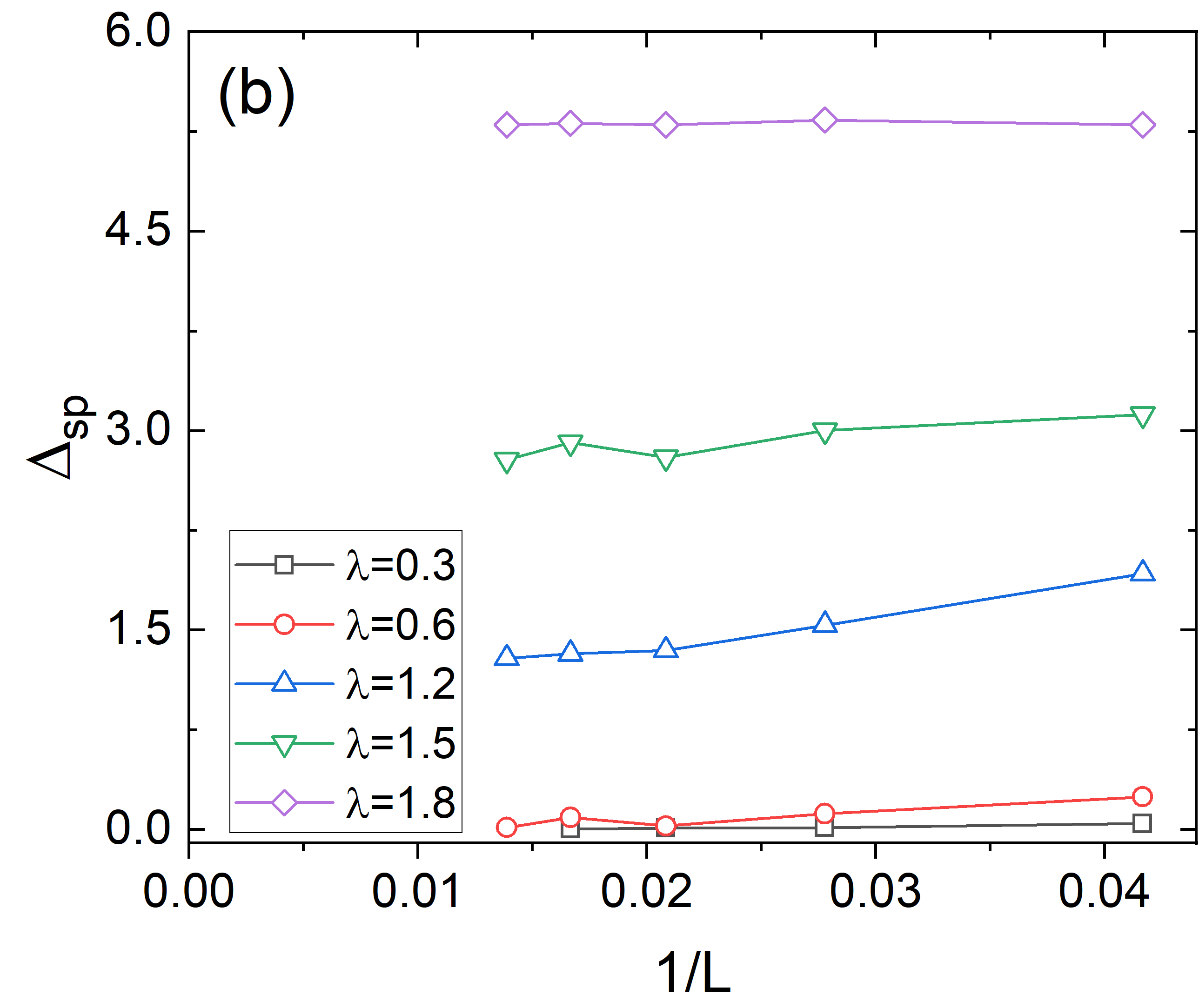}}
\caption{The results for single-particle gap $\Delta_{\mathrm{sp}}$ in the adiabatic limit $\omega_0=0$. (a) $\Delta_{\mathrm{sp}}$ as the function of $\lambda$. (b) A finite size extrapolation of $\Delta_{\mathrm{sp}}$ for some typical $\lambda$ in metallic disordered state ($\lambda=0.3$), metallic iVBS phase ($\lambda=0.6$), insulating iVBS phase ($\lambda=1.2,1.5$) and insulating sVBS phase ($\lambda=1.8$).  }
\label{FigS2}
\end{figure}

We also evaluate the single-particle gap for the ground state at adiabatic limit. The single-particle gap $\Delta_{\mathrm{sp}}$ can be obtained directly from the fermion spectrum under the optimal phonon configuration. \Fig{FigS2}(a) shows $\Delta_{\mathrm{sp}}$ as the function of coupling strength $\lambda$. $\Delta_{\mathrm{sp}}$ turns on for $\lambda>\lambda_{FS}\approx 1.0$. Namely, for $\lambda_{c1}<\lambda<\lambda_{FS}$, there remains a Fermi surface in iVBS phase. Meanwhile, for iVBS phase with $\lambda_{FS}<\lambda<\lambda_{c2}$, the single-particle spectrum is fully gapped despite the existence of a gapless Goldstone mode. The finite size behavior of $\Delta_{\mathrm{sp}}$ is exemplified in \Fig{FigS2}(b) by some typical $\lambda$ in either Fermi liquid, iVBS or sVBS phase. 

\subsection{C. Absence of symmetry-breaking in the quantum spin liquid phase}
\begin{figure}[t]
\subfigure{\label{figS3a}\includegraphics[width= 0.32\linewidth]{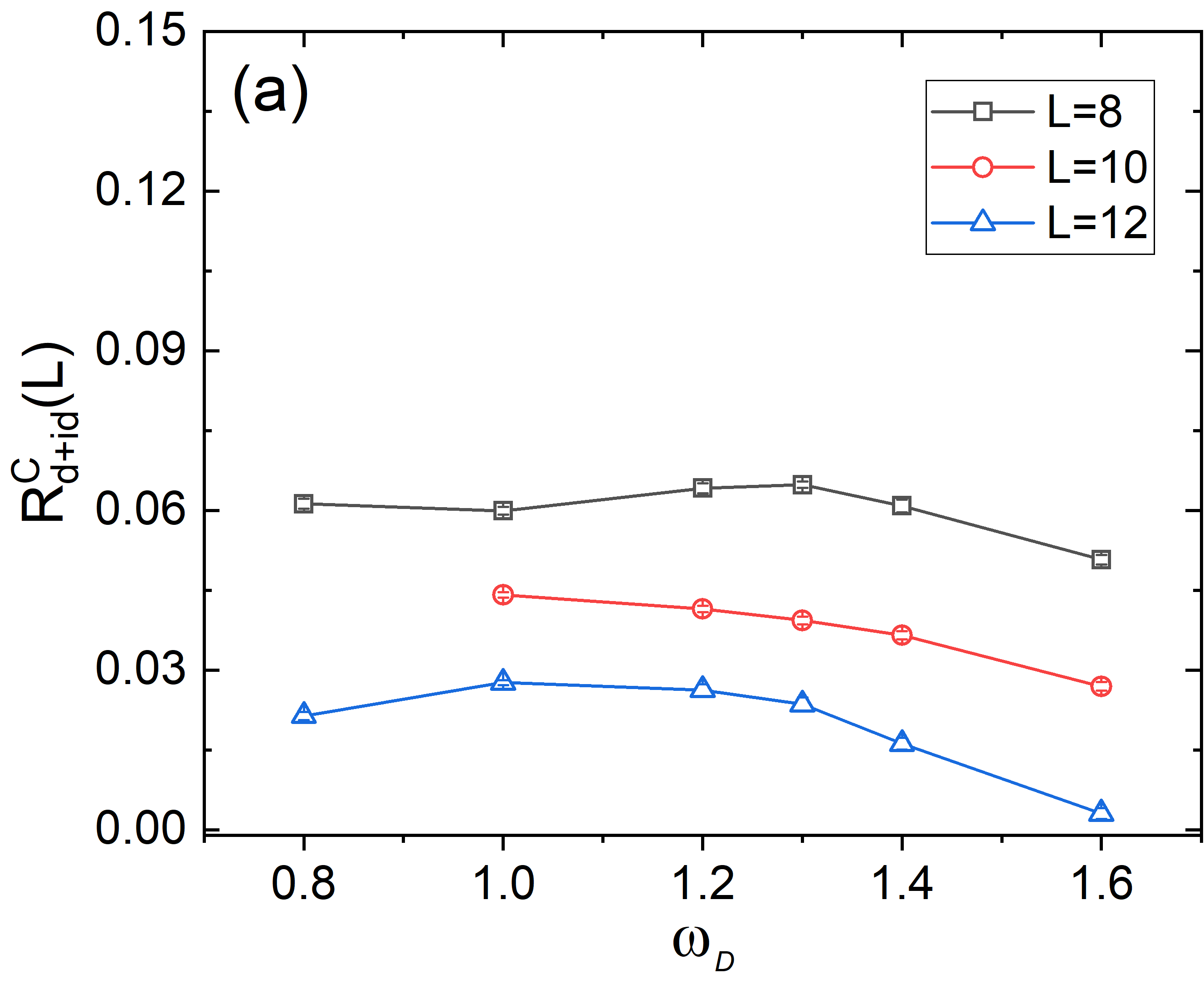}}~~
\subfigure{\label{figS3b}\includegraphics[width= 0.32\linewidth]{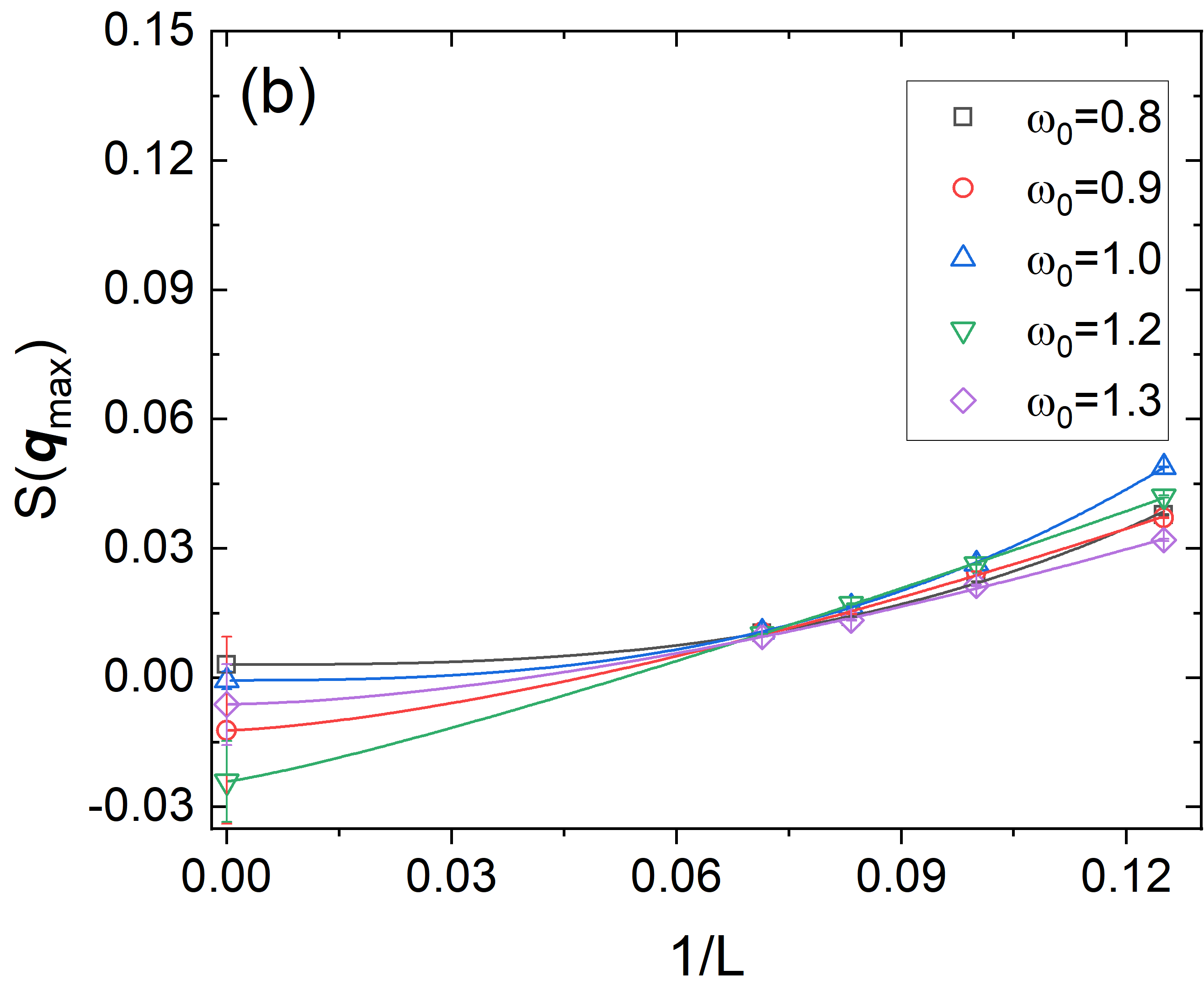}}~~
\subfigure{\label{figS3c}\includegraphics[width= 0.313\linewidth]{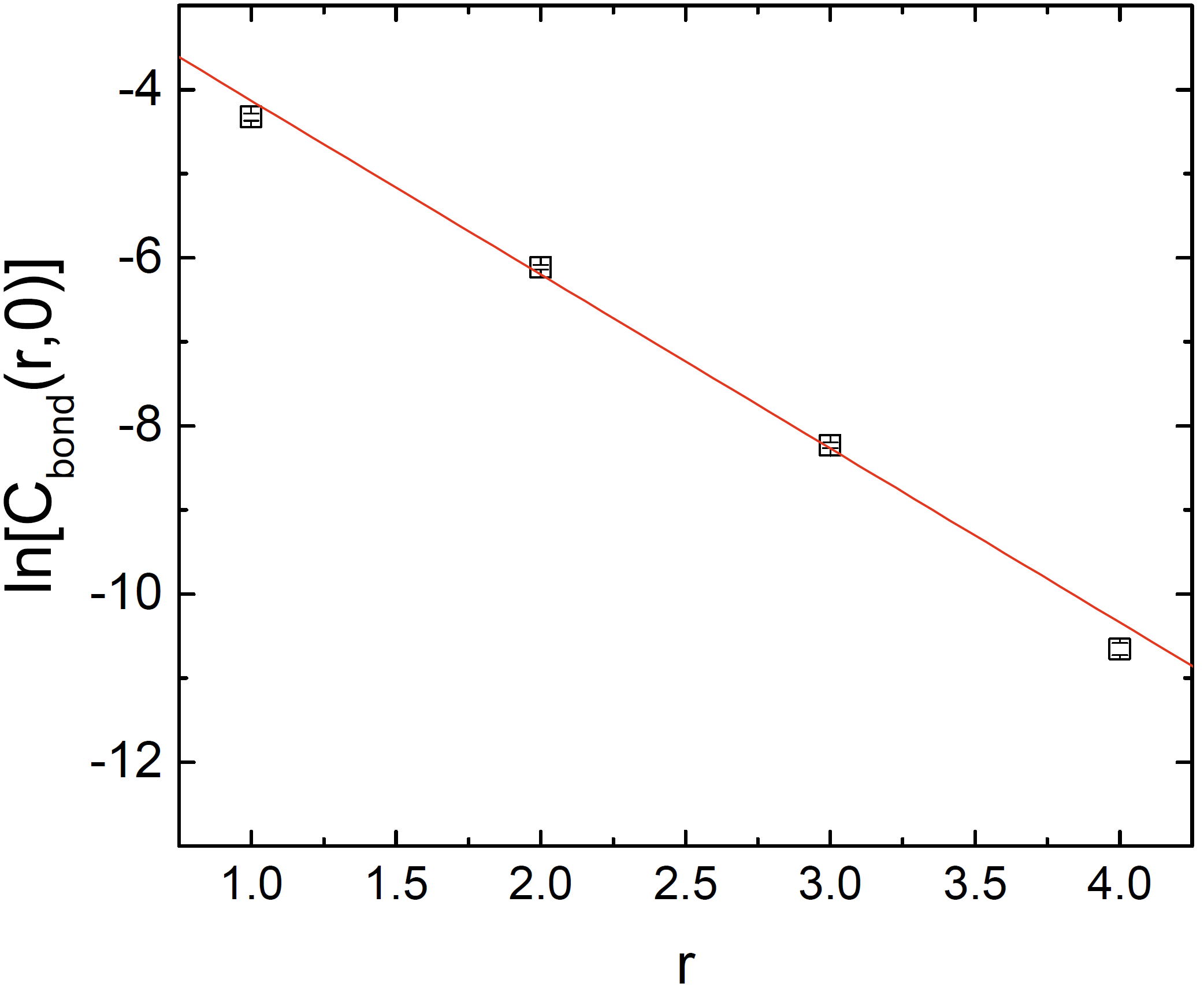}}
\caption{QMC results for various types of superconducting pairing and VBS orders. (a) Correlation-length ratio for $d+\imth d$ pairing as the function of $\omega_0$. (b) The finite-size scaling of the VBS structure factors at peaked momentum in the regime of QSL. (c) %{\color{blue} 
Logarithmic fitting of the bond-density ($\hat B_{ij}$) correlation function,  
$C_{\rm{bond}}[(r,0)]$ versus distance $r$  
for $\omega_0=1.0$ and $\lambda=2.16$ %. 
with system length $L=16$. The inverse of the slope yields 
the  estimate of the correlation length  $\xi=0.48(4)$.}
\label{FigS3}
\end{figure}

In this section, we demonstrate that the intermediate quantum spin liquid unveiled by the QMC simulation is absent from the long-range spontaneous symmetry-breaking order. 
In the main text, we have shown that the on-site superconducting order and staggered VBS order are short-range with no ordering. Here, we confirm the absence of other types of superconducting pairing and VBS long-range orders. In \Fig{FigS3}(a), we present the results of correlation-length ratios for the order parameter of $d+\imth d$ SC, which clearly confirm the nature of short-range $d+\imth d$ SC. For the VBS order, we consider $\sqrt{12}\times \sqrt{12}$ VBS ordering which exists in the triangular quantum dimer model. However, the structure factors of VBS order do not exhibit pronounced peak at the momentum corresponding to the $\sqrt{12}\times \sqrt{12}$ VBS ordering, which indicates the absence of $\sqrt{12}\times\sqrt{12}$ VBS long-range order. Additionally, we present the structure factors of VBS order at peaked momentum for different $L$ in the regime of QSL and perform the extrapolation of the results versus $\frac{1}{L}$, as depicted in \Fig{FigS3}(b). The extrapolated results vanish in the thermodynamic limit $L\rightarrow \infty$, which unambiguously confirm the absence of any long-range VBS order, including the $\sqrt{12}\times \sqrt{12}$ VBS order. %{\color{red} 
Furthermore, we evaluated the real-space correlation function of the bond operator $C_{\rm{bond}}[(r,0)]$
in the QSL regime % 
for parameters $\omega_0 = 1.0$ and $\lambda = 2.16$, where $(r,0)$ represents the bond-bond separation % 
along the horizontal direction % 
by distance $r$. The correlation length of the bond operators can be extracted from the slope of %  
the log of $C_{\rm{bond}}[(r,0)]$ versus distance $r$. The result of fitting is presented in \Fig{FigS3}(c), in which the inverse of the slope yields an %estimation of 
estimate of the bond correlation length $\xi =0.48(4)$.  Because the system size in our numerical simulation is much larger %compared with
than the bond-bond correlation length, we think the finite-size effects should be negligible and the conclusion of the absence of VBS long-range order in the QSL regime should thus be reliable.

\begin{figure}[b]
        \centering
        \includegraphics[width=0.8\linewidth]{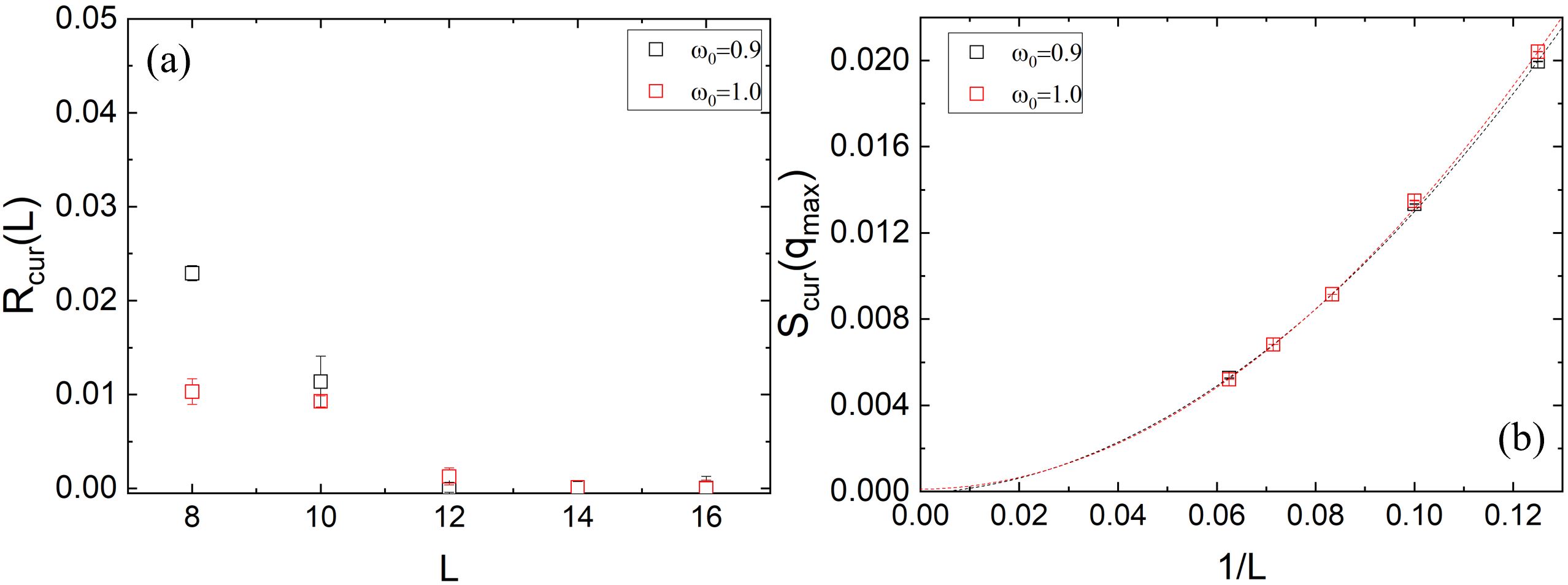}
        \caption{(a) The correlation-length ratio of loop-current operators versus $L$ for $\omega_0 = 0.9,1.0$ and $\lambda=2.16$. (b) The structure factor of  current order at peaked momentum versus $1/L$ for $\omega_0 = 0.9,1.0$ and $\lambda=2.16$. Second-order polynomial function of $1/L$ is used to extrapolate the results to thermodynamic limit $L \rightarrow \infty$.   }
        \label{FigS4}
\end{figure}
We further study other types of possible symmetry-breaking order, especially the loop current order which spontaneously breaks time-reversal symmetry. We computed the correlation function of the loop-current operator by QMC, and the results of correlation-length ratio and structure factors in the QSL regime are shown in \Fig{FigS4}. We fix $\lambda= 2.16$ and $\omega_0= 0.9$ and $1.0$, which reside in the QSL regime. The correlation-length ratios at peaked momentum decrease with system size, as depicted in \Fig{FigS4}(a), indicating that the current correlation is short-range. Moreover, the structure factors of loop-current order at peaked momentum vanish in the thermodynamic limit, as shown in \Fig{FigS4}(b), further confirming the absence of long-range loop current order.

\begin{figure}[tbp]
        \centering
        \includegraphics[width=0.7\linewidth]{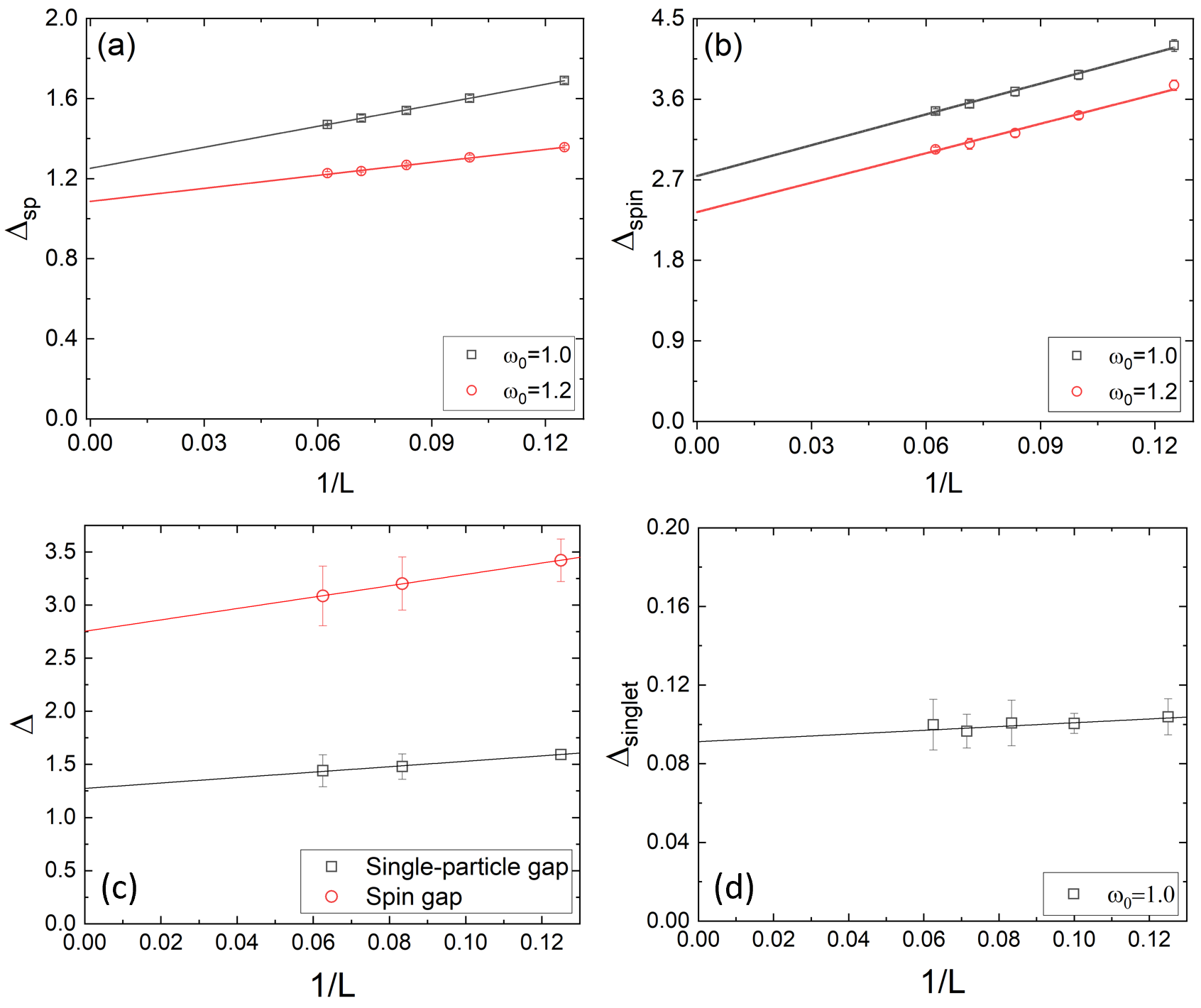}
        \caption{The QMC results of spectral gaps. Electron-phonon coupling strength is fixed as $\lambda=2.16$. (a) Single-particle gaps versus $1/L$ in the regime close to the QSL-SC transition. The intercept of linear fitting indicates the finite single-particle gap at thermodynamic limit. (b) Spin gaps versus $1/L$ in the regime close to the QSL-SC transition. The intercept of linear fitting indicates the finite spin gap at thermodynamic limit. (c) Single-particle and spin gaps versus $1/L$ in the staggered VBS phase with fixing $\omega_0=0.6$. The linear fitting of the results yields the finite single-particle and spin gaps at thermodynamic limit. (d) The results of spin singlet gap in the QSL regime with fixing $\omega_0=1.0$. The linear fitting of the results yields the finite spin singlet gap at thermodynamic limit.  }
        \label{FigS5}
\end{figure}

\subsection{D. Details of quantum Monte-Carlo simulations on spectral gaps}

In the main text, we presented the results of single-particle and spin gaps in the QSL regime. In this section, we provide more details of extracting spectral gap in QMC simulations. The single-particle gap is achieved by evaluating the imaginary-time correlator of single-particle  operators: $G_{\rm single}(\vec{k},\tau) = \langle\hat{c}_{\vec{k}}(0)\hat{c}^\dagger_{\vec{k}}(\tau)\rangle \propto e^{-\Delta_{\rm single}(\vec{k})\tau}$, where $\Delta_{\rm single}$ is the single-particle gap at momentum $\vec{k}$. We compute $\Delta_{\rm single}(\vec{k})$ with fixed linear system size $L$ at different $\vec{k}$, with the minimum value of $\Delta_{\rm single}(\vec{k})$ in momentum space defined as single-particle gap $\Delta_{\rm single}$ with system size $L$. We then perform linear fitting of single-particle gap $\Delta_{\rm single}$ versus $1/L$, with the intercept giving rise to the result at thermodynamic limit $L \rightarrow \infty$. Similarly, we compute spin gap via the imaginary-time correlator of spin operator $G_{\rm spin}(\vec{k},\tau) = \langle\hat{S}^+_{\vec{k}}(0)\hat{S}^-_{\vec{k}}(\tau)\rangle \propto e^{-\Delta_{\rm spin}(\vec{k})\tau}$. We implement the same procedure of finite-size scaling to determine the spin gap at thermodynamic limit. 
 
In addition to the results of single-particle and spin gaps in QSL regime, we provide the numerical results of spectral gaps in other phases in this section. The single-particle and spin gaps in the regime close to the SC-QSL transition are depicted in \Fig{FigS5}. The electron-phonon coupling strength is fixed at $\lambda=2.16$. With decreasing phonon frequency $\omega_0$, a continuous transition occurs from QSL to SC phase at $\omega_0 = 1.2$. As shown in \Fig{FigS5}(a) and (b), both single-particle and spin gaps are finite around the QSL-SC transition. We also evaluate the single-particle and spin gaps in the staggered VBS ordered phase with fixing $\lambda=2.16$ and $\omega_0=0.6$. The results indicate that both single-particle and spin gaps are finite in the staggered VBS ordered phase (shown in \Fig{FigS5}(c)). 

Furthermore, to further confirm the nature of the QSL phase, we compute the spin singlet gap in the QSL regime. The spin-singlet gap is achieved by evaluating the imaginary-time correlation function of hopping operator $\hat{B}_{ij} = \sum_{\sigma} (c^\dagger_{i\sigma}c_{j\sigma} + h.c.)$ on NN bond $\avg{ij}$. The results of spin-singlet gap in the QSL regime with fixed $\omega_0=1.0$ and $\lambda=2.16$ are presented in \Fig{FigS5} (d). The extrapolation to $L\rightarrow \infty$ demonstrates the finite spin-singlet gap in the thermodynamic limit, confirming that the QSL phase unveiled in our study is fully gaped.

\subsection{E. Critical property of phase transition between quantum spin liquid and superconductivity}

To determine the critical property of the phase transition between the QSL and SC phases, we evaluate the real-space correlation function of on-site superconducting pairing and extract the anomalous dimension of SC order at the transition. At the critical point of QSL-SC phase transition, the on-site SC correlation function displays algebraically decaying scaling behavior at sufficiently large distance: $C(\vec{R}) = \frac{1}{L^2}\sum_i \langle O_{\rm SC}(\vec{r}_i)O_{\rm SC}(\vec{r}_i+\vec{R})\rangle \propto \frac{1}{R^{1+\eta}}$, where $\hat{O}_{\rm SC}(\vec{r}_i) = c_{i\uparrow}c_{i\downarrow}$ is the on-site pairing operator at site $i$ and $\eta$ is anomalous dimension of the SC order parameter at the QSL-SC critical point. If the transition from gapped QSL to SC is driven by the condensation of fractionalized holon in QSL, which is the ``square root" of SC order parameter, the transition belongs to the XY$^{*}$ universality class. Compared with the conventional XY transition with anomalous dimension $\eta_{XY}\approx 0.03$, the XY$^{*}$ transition features a much larger anomalous dimension for the XY order parameter $\eta_{XY^*}\approx 1.49$~\cite{Isakov2012Science}.

\begin{figure}[tb]
        \centering
        \includegraphics[width=0.8\linewidth]{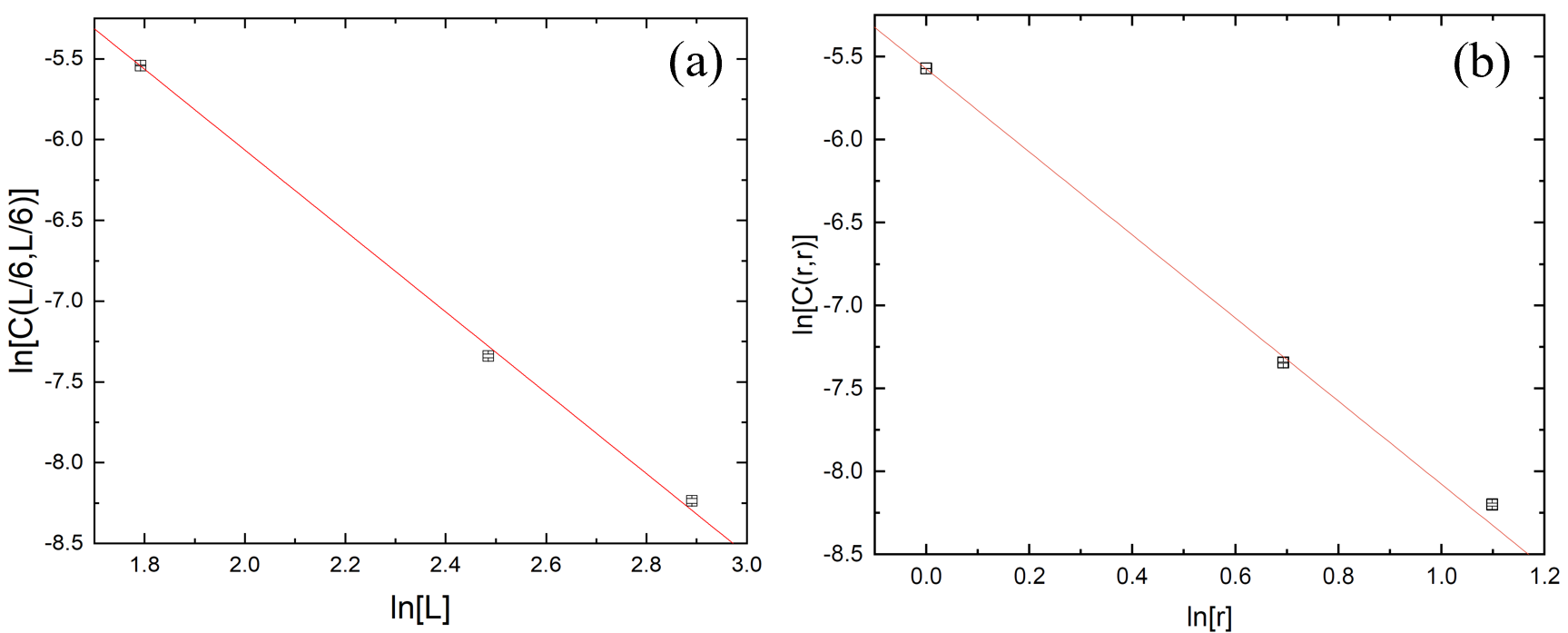}
        \caption{The results of correlation function of on-site pairing operators $C(\vec{R})$ at the QSL-SC transition point $\omega_0=1.2$ and $\lambda=2.16$. (a) The correlation function $C(\frac{L}{6},\frac{L}{6})$ versus linear system size $L$ for $L=6,12,18$. The slope of double-logarithmic fitting yields anomalous dimension $\eta = 1.45\pm 0.08$.  (b) The correlation function $C(r,r)$ versus $r$ with fixing $L=18$. The slope of double-logarithmic fitting yields anomalous dimension $\eta = 1.50\pm 0.09$.    }
        \label{FigS6}
\end{figure}

The QMC results of SC correlation function are presented in \Fig{FigS6}. In the system with periodic-boundary condition, the correlation function deviates from the scaling form of $\frac{1}{R^{1+\eta}}$ as the distance $R$ is close to half of linear system size, namely $\frac{L}{2}$. Hence, to access the relatively accurate value of anomalous dimension with the results of finite system size, 
one can perform the extrapolation of the correlation function with the distance away from $\frac{L}{2}$. For instance, in Ref. \cite{Isakov2012Science}, the authors implemented the scaling extrapolation of correlation function with $R = \frac{L}{6}$ versus $L$ and reached the value of anomalous dimension $\eta= 1.49(2)$. Here, we employ the similar procedure and extract the anomalous dimension from the correlation function of SC order parameter $C(\vec{R})$ with $\vec{R} = (\frac{L}{6},\frac{L}{6})$. As shown in \Fig{FigS6}(a), the extrapolation yields the anomalous dimension $\eta = 1.45 \pm 0.08$, which is consistent with the result of 3d XY$^*$ transition within fitting error. To further confirm the result, we performed the fitting procedure on the results of correlation function $C(r,r)$ for a fixed linear system size $L$ and $r<\frac{L}{5}$, which are shown in \Fig{FigS6} (b). For the case $L=18$, the linear fitting of $\ln[C(r,r)]$ versus $\ln(r)$ yields the results of anomalous dimension of SC order parameter $\eta=1.50\pm 0.09$, which is also consistent with the result of XY$^*$ universality class. The results obtained by different ways of fitting are consistent with each other within fitting error, further substantiating the conclusion that the QSL-SC transition belongs to the XY$^*$ universality class.

\subsection{F. Strong-coupling limit}

There are only three independent energy scales in this problem: $t$ (assuming $t>0$), $\omega_0$, and $\Veph \equiv g^2/K$, which characterizes the strength of the electron-phonon coupling. We emphasize that $\Veph$ is a free parameter that has no relation to $t$. We will consider the problem at half-filling (i.e., the number of electrons per site $n= 1$) in the strong coupling limit with $\Veph$ being the largest energy scale, namely $\Veph \gg t$ and $\Veph \gg \omega_0$.

\subsubsection{Low energy subspace: hard-core bond dimers}

We first obtain the low-energy subspace of the classical problem in the static limit, $\omega_0=0$ (i.e. $M=\infty$). In other words, we wish to identify the phonon configurations that minimize the adiabatic ground state energy, $E_0[\{ X_{\langle ij\rangle}\}]\equiv E_{\text{el},0}[\{ X_{\langle ij\rangle}\}]+\sum_{\langle ij \rangle}K X_{\langle ij \rangle}^2/2$, where the $E_{\text{el},0}[\{ X_{\langle ij\rangle}\}]$ is the ground state energy of the electronic system specified by the phonon configurations. For the half-filled case $n=1$, to the leading order 
of $t/\Veph$, we will show that there is an extensively degenerate ground state manifold that is in one-to-one correspondence with the states of maximal density hard-core dimers on the same lattice.

When $t=0$, the only electronic hopping between sites is phonon mediated, one can expect that the low energy states of the model in this limit consist of a collection of {\it disconnected} clusters of sites, with electrons (of each spin) occupying half number of states with lowest possible energy. One type of cluster that has optimal energy is a dimer that is localized on a bond with $X_{\langle ij\rangle} = \pm 2g/K$, which has the energy per electron $\epsilon =-\Veph$. Moreover, a tetramer state that is localized on a four-sided plaquette with phonon coordinates on the four sides $(X_1, X_2, X_3, X_4) = (a,b,a,-b)$ satisfying $a^2+b^2 = (2g/K)^2$ is also degenerate. Therefore, in this extreme limit of $\omega_0=0$ and $t=0$, an extensive family of states can be constructed corresponding to all allowed hard-core dimer and tetramer configurations on the lattice.

However, for finite (but small) $t$, the degeneracy under $X_{\langle ij\rangle} \to -X_{\langle ij\rangle}$ within a dimer and the degeneracy between dimer and tetramer are lifted to the leading order correction of $t$. Specifically, only dimers with $X_{\langle ij\rangle} =  2g/K$ have the optimal energy per electron $\epsilon=-\Veph -t$. Consequently, for the low energy states, only the hard-core dimer states with $X_{\langle ij\rangle} =  2g/K$ and electrons occupying the bonding orbital need to be considered for positive $t$.

We have not constructed an analytic proof that the dimer covering states constitute the ground states of the system. However, we have performed extensive numerical investigation (by gradient descent optimization of the total energy) in the $\omega_0=0$, $t\ll \Veph$ case at half-filling on the triangular lattice. Depending on the initially randomly assigned lattice configuration, we end up in different local minima of the energy function $E_0$. While in all cases the phonon configuration $\{X_{\langle ij \rangle }\}$ we find are close-packed patterns of hard-core dimers. We thus conclude that these are all the degenerate ground-states of the model at half-filling to the leading order effect of $t$.

The above discussions have not considered the spread of phonon wavefunctions of the states in the low-energy subspace. We will revisit this issue later in the next section.

\subsubsection{Effective dimer interactions}

There are two contributions to the effective dimer interactions: hopping-mediated repulsion (second order in $t$), and an attraction caused by the zero point energy (linear order in $\omega_0$).

We first see the effect of $t$. To the second order in $t$, there can be three types of virtual processes involving pairs of neighboring dimers, which will give rise to effective dimer repulsions.

First, the processes can involve two uncovered bonds, one site in one of the dimers, and two sites in the other dimer (possible on triangular and kagome lattices),
\begin{align}
\begin{tikzpicture}[baseline={([yshift=-.5ex]current bounding box.center)}]
    \draw coordinate (a) at (0.0,0.0);
    \draw coordinate (b) at (0.6,0.0);
    \draw coordinate (c) at (0.3,0.45);
    \draw coordinate (d) at (0.9,0.45);
    \draw[color=black!60] (a) circle (0.06);
    \draw[color=black!60] (b) circle (0.06);
    \draw[color=black!60] (c) circle (0.06);
    \draw[color=black!60] (d) circle (0.06);
    \draw[black, ultra thick] (a)--(b);
    \draw[black, ultra thick] (c)--(d);
    \draw[-stealth,black, thin] (0.05,0.07)--(0.25,0.4);
    \draw[-stealth,black, thin] (0.35,0.35)--(0.55,0.07);
\end{tikzpicture}
\end{align}
However, these processes will have zero overall effect, since the different sequences of the virtual hoppings will result in a cancellation.

Second, the processes can only involve one uncovered bond and one site in each of the two dimers (possible on all lattices),
\begin{align}
\begin{tikzpicture}[baseline={([yshift=-.5ex]current bounding box.center)}]
    \draw coordinate (a) at (-0.4,0.4);
    \draw coordinate (b) at (0,0);
    \draw coordinate (c) at (0.4,0.4);
    \draw coordinate (d) at (0.8,0);
    \draw[color=black!60] (a) circle (0.06);
    \draw[color=black!60] (b) circle (0.06);
    \draw[color=black!60] (c) circle (0.06);
    \draw[color=black!60] (d) circle (0.06);
    \draw[black, ultra thick] (c)--(d);
    \draw[black, ultra thick] (a)--(b);
    \draw[stealth-stealth,black, thin] (0.05,0.05)--(0.35,0.35);
\end{tikzpicture}
\end{align}
whose effect on the ground state energy only depends on the number of uncovered bonds (thus the number of dimers), and only gives rise to a configuration-independent constant shift in total energy. Just to be complete, assuming no other second-order processes present, the energy on each uncovered bond $\langle ij \rangle$ is $E = - \frac{(t+g X_{\langle ij \rangle})^2}{4\Veph} + \frac{KX^2_{\langle ij \rangle}}{2}$, which can be minimized to $-\frac{t^2}{2\Veph}$ by taking $X_{\langle ij \rangle} = t/g$.

Third, there can be processes that involve two uncovered bonds and two sites in each of the two dimers (only possible on square and triangular lattices),
\begin{align}
\begin{tikzpicture}[baseline={([yshift=-.5ex]current bounding box.center)}]
    \draw coordinate (a) at (0.0,0.0);
    \draw coordinate (b) at (0.6,0.0);
    \draw coordinate (c) at (0.3,0.45);
    \draw coordinate (d) at (0.9,0.45);
    \draw[color=black!60] (a) circle (0.06);
    \draw[color=black!60] (b) circle (0.06);
    \draw[color=black!60] (c) circle (0.06);
    \draw[color=black!60] (d) circle (0.06);
    \draw[black, ultra thick] (a)--(b);
    \draw[black, ultra thick] (c)--(d);
    \draw[-stealth,black, thin] (0.05,0.07)--(0.25,0.4);
    \draw[-stealth,black, thin] (0.85,0.35)--(0.65,0.07);
\end{tikzpicture}
\end{align}
Denoting the phonon coordinates on the uncovered bonds as $X$ and $Y$, and taking into account of all second-order processes, we obtain the expression for the energy on the two uncovered bonds:
\begin{align}
    E = -\frac{(g X + t)^2}{4\Veph} -\frac{(g Y + t)^2}{4\Veph} + \frac{(g X +t)(g Y +t)}{2\Veph}+ \frac{KX^2}{2}+ \frac{KY^2}{2}
\end{align}
Note that, there is an extra minus sign for the 
third term above, resulting from the intermediate anti-bonding state. This energy can be minimized to $0$ by taking $X=Y=0$. Comparing this with that in the previous case, we obtain the effective repulsion
\begin{align}
    \hat{V} = V \sum_{\plaquette}\left[ \left|\plaquettedimerone
    \right\rangle \left\langle \plaquettedimerone \right| + \left|\plaquettedimertwo
    \right\rangle \left\langle \plaquettedimertwo \right| \right]
\end{align}
with $V = t^2/\Veph>0$ for square and triangular lattices.

Next, we analyze the effect of zero point energy (ZPE) at finite but very small small $\omega_0$ (large ion mass $M$). This amounts to the analysis of the stability matrix of the system, which can be calculated by performing a second-order perturbation theory in virtual perturbations $\delta X_{\langle ij\rangle}$. The analysis is very similar to the above one, which we won't repeat. The main conclusion is the following: 

For each covered bond, the stiffness $K$ and thus the ZPE$=\omega_0/2$ is unchanged, and the phonon wavefunction has Gaussian spread $x_0 = 1/\sqrt[4]{MK}$. For each uncovered bond, if it does not belong to a plaquette with two dimers, its stiffness is reduced from $K \rightarrow K/2$, thus ZPE is reduced from $\omega_0/2 \rightarrow \sqrt{2}\omega_0/4$, and Gaussian spread of the phonon wavefunction increases to $x_0' = 1/\sqrt[4]{MK/2}$. Lastly, for a plaquette with two dimers, if we denote the phonon coordinates on the uncovered bonds as $X$ and $Y$, then the classical energy can be perturbatively evaluated as:
\begin{align}
    E = K(\delta X+\delta Y)^2/4
\end{align}
which implies that the normal mode $X_+\equiv (\delta X+\delta Y)/\sqrt{2}$ has stiffness $K$, ZPE $\omega_0/2$, and Gaussian spread $x_0=1/\sqrt[4]{MK}$, but the normal mode $X_-\equiv (\delta X-\delta Y)/\sqrt{2}$ has zero stiffness and ZPE, and an extended wavefunction (with a large spread width $\sim g/K$, as expected based on the existence of a degenerate manifold of tetramer states in the $t=0$ limit).  Effectively, the difference in ZPE of different dimer configurations induces a dimer attraction $V=-\frac{\sqrt{2}-1}{2}\omega_0$. Consequently, to the second order of $t$ and first order of $\omega_0$, the potential energy of two dimers on the same plaquette is $V=t^2/\Veph -\frac{\sqrt{2}-1}{2}\omega_0$.

\subsubsection{An effective model: quantum dimer model}

Now we consider the effect of a finite (but small) $\omega_0$ on dimer kinetics or resonance. In this case, the phonon dynamics enables the phonon configurations to tunnel between nearby minima in the energy landscape. In terms of the effective dimer degrees of freedom, this results in the resonance of dimer configurations. The minimal resonance is
\begin{align}
    \hat{J} = -J \sum_{\plaquette}\left[ \left|\plaquettedimerone
    \right\rangle \left\langle \plaquettedimertwo \right| + \left|\plaquettedimerone
    \right\rangle \left\langle \plaquettedimertwo \right| \right]
\end{align}
where $J$ should be an increasing function of $\omega_0$, and vanish in the static limit and is necessarily vanish to all orders in $t$ in the $\omega_0 \to 0$ limit as they are tunnelling processes - akin to small bipolaron hopping - involving rearrangements of the phonon coordinates. A crude estimate can be made by noting that the barrier height is $\sim g^2/K $ and the distance tunneled is $\sim g/K$, meaning that $J \sim \omega_0 \exp[ - \alpha g^2/(K\omega_0)]$ where $\alpha $ is a number of order one.

Combining both the effective dimer repulsion generated by small $t$ and the effective dimer resonance generated by small $\omega_0$, we reach an effective model in the strong coupling, adiabatic limit of the problem, which is the quantum dimer model
\begin{align}
\hat{H}_{\text{dimer}} = \hat{V}+ \hat{J}.
\end{align}

\subsubsection{Review on the phase diagram of the quantum dimer model}

The $T=0$ phase diagram of the quantum dimer model on the triangular lattice has been obtained~\cite{Moessner2001PRL,PhysRevB.71.224109}, which we briefly summarize now. First of all, $J=V$ is a special point (RK point~\cite{Kivelson1988PRL,moessner2010quantum-misc}) whose ground state is an equal amplitude superposition of all dimer configurations in a given topological sector, i.e. a short-ranged RVB state. This RVB state exhibits $\mathbb{Z}_2$ topological order on the triangular lattice, and is stable in a range of $\nu_c J <V<J$ with $\nu_c \lesssim 0.8$; two different VBSs occur for other ranges of parameters: staggered (for $V>J>0$) and $\sqrt{12}\times \sqrt{12}$ (for $0<V<\nu_c J$). Comparing with the numerical results in the strong coupling regime, we find the phase in the adiabatic limit (small $\omega_0$ and thus small $J$) is indeed a staggered VBS, whereas when $\omega_0$ is moderate a quantum spin liquid phase emerges, which likely has a $\mathbb{Z}_2$ topological order.

\end{appendices}

\end{document}